\def\IDvalue{PE}
\def\DIRvalue{Pestun}
\def\titlevalue{Review of localization in geometry}
\def\authorvalue{Vasily Pestun}
\def\shortauthorvalue{\authorvalue}
\def\addressvalue{Institut des Hautes \'Etudes Scientifique, France\\
\tt pestun@ihes.fr}
\def\abstractvalue{Review of localization in geometry: equivariant
  cohomology, characteristic
  classes, Atiyah-Bott formula, Atiyah-Singer equivariant index formula,
Mathai-Quillen formalism}
\def\preprintvalue{}

\ifx\ifLONG\undefined
\documentclass[12pt,psamsfonts,reqno]{amsart} 

\newcommand{\chapterauthor}[1]{
\begin{center}
{\bf \normalsize  #1}
\end{center}
}

\newcommand{\chapteraddress}[1]{
\begin{center}
{ \small \it \addressvalue}
\end{center}
}

\newcommand{\chapterabstract}[1]{
\vspace{\baselineskip}
\begin{center}
\textbf{\small Abstract}
\end{center}
#1}

\newcommand{\chapterheader}{

\chapter[\titlevalue{}  (by \shortauthorvalue)]{\titlevalue}
\label{Chapter\IDvalue}
\chapterauthor{\authorvalue}
\chapteraddress{\addressvalue}
\chapterabstract{\abstractvalue}
\tightmtctrue
\minitoc
}

\newcommand{\documentheader}{
\begin{flushright} \small
  \preprintvalue
 \end{flushright}

\begin{center}
{\bf \Large \titlevalue}
\end{center}

\chapterauthor{\authorvalue}
\chapteraddress{\addressvalue}
\chapterabstract{\abstractvalue}

\medskip

This is a contribution to the review volume ``Localization techniques
in quantum field theories'' (eds. V.~Pestun and M.~Zabzine) which
contains 17 Chapters available at \cite{ContributionSummary}

\tableofcontents
}

\newcommand{\ifvolume}[2]{\ifx\ifLONG\undefined#2\else#1\fi}

\newcommand{\documentfinish}{
\ifx\ifLONG\undefined
\bibliographystyle{bibreview} 
\bibliography{\IDvalue,review}  
\end{document}
\else
\addcontentsline{toc}{section}{References}
\providecommand{\href}[2]{#2}\begingroup\raggedright\begin{thebibliography}{10}

\bibitem{ContributionSummary}
V.~Pestun and M.~Zabzine, eds., {\em Localization techniques in quantum field
  theory}, vol.~xx.
\newblock Journal of Physics A, 2016.
\newblock \href{http://arxiv.org/abs/1608.02952}{{\tt 1608.02952}}.
\newblock \url{https://arxiv.org/src/1608.02952/anc/LocQFT.pdf},
  \url{http://pestun.ihes.fr/pages/LocalizationReview/LocQFT.pdf}.

\bibitem{PECartan1951}
H.~Cartan, ``Notions d'alg\`ebre diff\'erentielle; application aux groupes de
  {L}ie et aux vari\'et\'es o\`u op\`ere un groupe de {L}ie,'' {\em Colloque de
  topologie (espaces fibr\'es), {B}ruxelles, 1950} (1951)  15--27.

\bibitem{PECartan1951a}
H.~Cartan, ``La transgression dans un groupe de {L}ie et dans un espace fibr\'e
  principal,'' {\em Colloque de topologie (espaces fibr\'es), {B}ruxelles,
  1950} (1951)  57--71.

\bibitem{PEGuillemin1999}
V.~W. Guillemin and S.~Sternberg,
  \href{http://dx.doi.org/10.1007/978-3-662-03992-2}{{\em Supersymmetry and
  equivariant de {R}ham theory}}.
\newblock Mathematics Past and Present. Springer-Verlag, Berlin, 1999.
\newblock \url{http://dx.doi.org/10.1007/978-3-662-03992-2}.
\newblock With an appendix containing two reprints by Henri Cartan [ MR0042426
  (13,107e); MR0042427 (13,107f)].

\bibitem{PEBerline:2004}
N.~Berline, E.~Getzler, and M.~Vergne, {\em Heat kernels and {D}irac
  operators}.
\newblock Grundlehren Text Editions. Springer-Verlag, Berlin, 2004.
\newblock Corrected reprint of the 1992 original.

\bibitem{PESzabo:1996md}
R.~J. Szabo, ``{Equivariant localization of path integrals},''
\href{http://arxiv.org/abs/hep-th/9608068}{{\tt arXiv:hep-th/9608068
  [hep-th]}}.

\bibitem{PECordes:1994sd}
S.~Cordes, G.~W. Moore, and S.~Ramgoolam, ``Large {N 2-D} {Y}ang-{M}ills theory
  and topological string theory,'' {\em Commun. Math. Phys.} {\bf 185} (1997)
  543--619,
\href{http://arxiv.org/abs/hep-th/9402107}{{\tt hep-th/9402107}}.

\bibitem{PEVergne:2006}
M.~Vergne, ``Applications of Equivariant Cohomology,''
  \href{http://arxiv.org/abs/math/0607389}{{\tt math/0607389}}.

\bibitem{PEMR0077122}
J.~Milnor, ``Construction of universal bundles. {I},'' {\em Ann. of Math. (2)}
  {\bf 63} (1956)  272--284.

\bibitem{PEMR0077932}
J.~Milnor, ``Construction of universal bundles. {II},'' {\em Ann. of Math. (2)}
  {\bf 63} (1956)  430--436.

\bibitem{PEKanno:1988wm}
H.~Kanno, ``{Weyl Algebra Structure and Geometrical Meaning of {BRST}
  Transformation in Topological Quantum Field Theory},''
\href{http://dx.doi.org/10.1007/BF01506544}{{\em Z. Phys.} {\bf C43} (1989)
  477}.

\bibitem{PELabastida:1988qb}
J.~M.~F. Labastida and M.~Pernici, ``{A Gauge Invariant Action in Topological
  Quantum Field Theory},''
\href{http://dx.doi.org/10.1016/0370-2693(88)91235-X}{{\em Phys. Lett.} {\bf
  B212} (1988)  56}.

\bibitem{PEPestun:2007rz}
V.~Pestun, ``{Localization of gauge theory on a four-sphere and supersymmetric
  Wilson loops},'' \href{http://dx.doi.org/10.1007/s00220-012-1485-0}{{\em
  Commun.Math.Phys.} {\bf 313} (2012)  71--129},
\href{http://arxiv.org/abs/0712.2824}{{\tt arXiv:0712.2824 [hep-th]}}.

\bibitem{PEBott:2001}
R.~Bott and L.~W. Tu, {\em Equivariant characteristic classes in the {C}artan
  model}.
\newblock World Sci. Publ., River Edge, NJ, 2001.

\bibitem{PEMR1288997}
N.~Berline and M.~Vergne, ``The equivariant {C}hern character and index of
  {$G$}-invariant operators. {L}ectures at {CIME}, {V}enise 1992,''.

\bibitem{PEBott:book}
R.~Bott and L.~W. Tu, {\em Differential forms in algebraic topology}, vol.~82
  of {\em Graduate Texts in Mathematics}.
\newblock Springer-Verlag, New York-Berlin, 1982.

\bibitem{PEMathaiQuillen}
V.~Mathai and D.~Quillen, ``Superconnections, {T}hom classes, and equivariant
  differential forms,''
  \href{http://dx.doi.org/10.1016/0040-9383(86)90007-8}{{\em Topology} {\bf 25}
  (1986) no.~1, 85--110}. \url{http://dx.doi.org/10.1016/0040-9383(86)90007-8}.

\bibitem{PEMR1094734}
M.~F. Atiyah and L.~Jeffrey, ``Topological {L}agrangians and cohomology,'' {\em
  J. Geom. Phys.} {\bf 7} (1990) no.~1, 119--136.

\bibitem{PEWu:2005pr}
S.~Wu, ``{Mathai-Quillen formalism},''
\href{http://arxiv.org/abs/hep-th/0505003}{{\tt arXiv:hep-th/0505003
  [hep-th]}}.

\bibitem{PEAtiyah1984}
M.~F. Atiyah and R.~Bott, ``The moment map and equivariant cohomology,'' {\em
  Topology} {\bf 23} (1984) no.~1, 1--28.

\bibitem{PEMR2069175}
A.~A. Kirillov, \href{http://dx.doi.org/10.1090/gsm/064}{{\em Lectures on the
  orbit method}}, vol.~64 of {\em Graduate Studies in Mathematics}.
\newblock American Mathematical Society, Providence, RI, 2004.
\newblock \url{http://dx.doi.org/10.1090/gsm/064}.

\bibitem{PEMR0294568}
B.~Kostant, ``Quantization and unitary representations. {I}.
  {P}requantization,''.

\bibitem{PEMR1701415}
A.~A. Kirillov, ``Merits and demerits of the orbit method,''
  \href{http://dx.doi.org/10.1090/S0273-0979-99-00849-6}{{\em Bull. Amer. Math.
  Soc. (N.S.)} {\bf 36} (1999) no.~4, 433--488}.
  \url{http://dx.doi.org/10.1090/S0273-0979-99-00849-6}.

\bibitem{PEAlekseev:1988ce}
A.~Alekseev and S.~L. Shatashvili, ``{Path Integral Quantization of the
  Coadjoint Orbits of the Virasoro Group and 2D Gravity},''
\href{http://dx.doi.org/10.1016/0550-3213(89)90130-2}{{\em Nucl. Phys.} {\bf
  B323} (1989)  719}.

\bibitem{PEAtiyah1974}
M.~F. Atiyah, {\em Elliptic operators and compact groups}.
\newblock Springer-Verlag, Berlin, 1974.
\newblock Lecture Notes in Mathematics, Vol. 401.

\bibitem{ContributionPZ}
V.~Pestun and M.~Zabzine, ``Introduction to localization in quantum field
  theory,'' {\em Journal of Physics A} {\bf xx} (2016)  000,
  \href{http://arxiv.org/abs/1608.02953}{{\tt 1608.02953}}.

\bibitem{PEWitten:1990bs}
E.~Witten, ``{Introduction to cohomological field theories},''
\href{http://dx.doi.org/10.1142/S0217751X91001350}{{\em Int.J.Mod.Phys.} {\bf
  A6} (1991)  2775--2792}.

\bibitem{PEWitten:1988ze}
E.~Witten, ``{Topological Quantum Field Theory},''
\href{http://dx.doi.org/10.1007/BF01223371}{{\em Commun.Math.Phys.} {\bf 117}
  (1988)  353}.

\bibitem{PENekrasov:2002qd}
N.~A. Nekrasov, ``{Seiberg-Witten prepotential from instanton counting},'' {\em
  Adv.Theor.Math.Phys.} {\bf 7} (2004)  831--864,
  \href{http://arxiv.org/abs/hep-th/0206161}{{\tt arXiv:hep-th/0206161
  [hep-th]}}.
To Arkady Vainshtein on his 60th anniversary.

\bibitem{ContributionHO}
K.~Hosomichi, ``$\mathcal{N}=2$ SUSY gauge theories on $S^4$,'' {\em Journal of
  Physics A} {\bf xx} (2016)  000, \href{http://arxiv.org/abs/1608.02962}{{\tt
  1608.02962}}.

\end{thebibliography}\endgroup

\fi
}

\newcommand{\documentfinishBBL}{
\addcontentsline{toc}{section}{References}
\ifx\ifLONG\undefined
\input{\IDvalue.separate.bbl}
\end{document}
\else
\input{\DIRvalue/\IDvalue.volume.bbl}
\fi
}

\ifx\ifLONG\undefined
\def\volcite#1{Contribution \cite{Contribution#1}}
\else 
\def\volcite#1{Chapter \ref{Chapter#1}}
\fi

\usepackage{ytableau}
\usepackage[margin=1in]{geometry}
\usepackage{amsmath,amssymb,amsfonts,mathtools}
\usepackage{amsxtra}
\usepackage{amscd}
\usepackage{xcolor}

 \usepackage{tikz-cd}
 \usetikzlibrary{matrix,arrows,decorations.pathmorphing}
 \usepackage[mathscr]{euscript}
 \usepackage{amscd}
 \usepackage{slashed}
 \usepackage{listings}
 \usepackage{graphicx}
 \usepackage{todonotes}

\usepackage[pagebackref=false]{hyperref}

 \definecolor{darkblue}{RGB}{32,32,156}
 \definecolor{darkgreen}{RGB}{32,128,32}
\hypersetup{colorlinks = true,extension = notused,linkcolor = darkblue,anchorcolor = red,citecolor = darkgreen,filecolor = red,pagecolor = red,urlcolor = darkblue}

 \usepackage{iftex}
 \ifxetex
         \usepackage{fontspec}
 \else
 \fi

\newcommand {\la} {\left \langle}
\newcommand {\ra} {\right \rangle}

\newcommand {\CalO} {\mathcal O}

\newcommand {\BR}   {\mathbb R}
\newcommand {\BZ}   {\mathbb Z}
\newcommand {\BC}   {\mathbb C}

\newcommand {\CP}   {\mathbb C \mathbb P}

 \newcommand {\ep}  {\epsilon}

\newcommand{\g}{\mathfrak{g}}

\DeclareMathOperator{\ch}{ch}
\DeclareMathOperator{\td}{td}

\DeclareMathOperator{\coker}{coker}

\DeclareMathOperator{\Hom}{Hom}

\DeclareMathOperator{\tr} {tr}

\DeclareMathOperator{\str} {str}
\DeclareMathOperator{\rk} {rk}

\DeclareMathOperator{\ad} {ad}
\DeclareMathOperator{\ind} {ind}

\DeclareMathOperator{\rank}{rank}

\newcommand{\SL}{SL}

\numberwithin{equation}{section}

\begin{document}

\documentheader
\else 

\chapterheader 

\fi

Foundations of equivariant de Rham theory have been laid in two papers
by Henri Cartan  \cite{PECartan1951} \cite{PECartan1951a}. The book by Guillemin and Sternberg \cite{PEGuillemin1999} covers Cartan's
papers and treats equivariant de Rham theory from the perspective
of  supersymmetry. See also the book by Berline-Vergne \cite{PEBerline:2004},
the lectures by Szabo \cite{PESzabo:1996md} and by Cordes-Moore-Ramgoolam
\cite{PECordes:1994sd}, and Vergne's review \cite{PEVergne:2006}.

\section{Equivariant cohomology}

\newcommand{\Lie}{\mathrm{Lie}}
\newcommand{\euler}{\mathrm{e}}

\newcommand{\Pf}{\mathrm{Pf}}

Let $G$ be a compact connected Lie group. Let $X$ be a $G$-manifold, which means that
there is a defined action $G \times X \to X$ of the group $G$ on the manifold
$X$. 

If $G$ acts freely on $X$ (all stabilizers are trivial) then the
space $X/G$ is an ordinary manifold on which the usual cohomology theory
$H^{\bullet}(X/G)$ is defined. If the $G$ action on $X$ is free, the $G$-equivariant cohomology groups $H_{G}^{\bullet}(X)$
are defined to be the ordinary cohomology $H^{\bullet}(X/G)$.

If the $G$ action on $X$ is not free, the naive definition of the
equivariant cohomology $H_{G}^{\bullet}(X)$ fails because $X/G$ is not an ordinary manifold. 
 If non-trivial stabilizers exist, the corresponding points on $X/G$ are not ordinary points but
fractional or stacky points. 

A proper topological definion of the \emph{$G$-equivariant cohomology}
$H_{G}(X)$ sets
\begin{equation}
  H_{G}^\bullet(X) = H^\bullet( X \times_{G} EG) = H^\bullet( (X \times EG)/G))
\end{equation}
  where the space $EG$, called
\emph{universal bundle}  \cite{PEMR0077122,PEMR0077932}  is a topological space
associated to $G$ with the following properties 
\begin{enumerate}
\item The space $EG$ is contractible 
\item The group $G$ acts freely on $EG$
\end{enumerate}

Because of the property (1) the cohomology theory of $X$ is isomorphic
to the cohomology theory of $X \times
EG$, and because of the property (2) the group $G$ acts freely on $X
\times EG$ and hence the quotient space $(X \times_{G} EG)$ has a well-defined ordinary cohomology theory. 

\section{Classifying space and characteristic classes}
If $X$ is a point $pt$, the ordinary cohomology theory $H^\bullet(pt)$ is elementary
\begin{equation}
H^{n}(pt,\BR) =   \begin{cases} \BR, \qquad n = 0 \\
    0, \qquad n > 0 
\end{cases}
\end{equation}
but the equivariant cohomology $H_{G}^\bullet(pt)$ is less trivial. Indeed, 
\begin{equation}
  H_{G}^{\bullet}(pt) = H^{\bullet}(EG/G) = H^{\bullet}(BG)
\end{equation}
where the quotient space $BG = EG/G$ is called \emph{classifying
  space}. 

The  terminology \emph{universal bundle} $EG$ and
\emph{classifying space} $BG$ comes from the fact that  any smooth
principal $G$-bundle on a manifold $X$ can be induced by a pullback $f^{*}$ of
the universal principal $G$-bundle $EG \to BG$ using a suitable smooth map $f: X \to
BG$. 

The cohomology groups of $BG$ are used to construct \emph{characteristic
  classes} of principal $G$-bundles.

 Let $\g = Lie(G)$ be the
real Lie algebra of a compact connected Lie group $G$. Let $\BR[\g]$ be the
space of real valued polynomial functions on $\g$, and let $\BR[\g]^{G}$
be the subspace of $\mathrm{Ad}_{G}$ invariant polynomials on $\g$.

For a principal $G$-bundle over a base manifold $X$ the Chern-Weil morphism 
\begin{equation}
\label{PEeq:CW}
  \begin{aligned}
  \BR[\g]^{G} &\to H^{\bullet}(X, \BR)\\
    p &\mapsto p(F_{A})
  \end{aligned}
\end{equation}
sends an adjoint invariant polynomial $p$ on the Lie algebra $\g$ to a
cohomology class  $[p(F_{A})]$ in $H^\bullet(X)$ where
$F_{A} = \nabla_{A}^2$ is the curvature 2-form of any connection $\nabla_{A}$
on the $G$-bundle. The cohomology class $[p(F_A)]$ does not depend on
 the choice of the connection $A$ and is called the \emph{characteristic
  class} of the $G$-bundle associated to the polynomial $p \in \BR[\g]^{G}$. 

The main theorem of Chern-Weil theory is that  the ring of characteristic classes $\BR[\g]^{G}$ 
  is isomorphic to the cohomology ring $H^\bullet(BG)$ of the classifying space
  $BG$:  the Chern-Weil morphism (\ref{PEeq:CW}) is an isomorphism
\begin{equation}
  \BR[\g]^{G} \stackrel{\sim}{\to} H^{\bullet}(BG, \BR)
\end{equation}

For the circle group  $G = S^1 \simeq U(1)$ the universal bundle $ES^1$ and
classifying space $BS^1$ can be
modelled as
\begin{equation}
ES^1 \simeq S^{2n+1}, \qquad BS^1 \simeq \CP^{n} \qquad\text{at}\qquad n \to \infty
\end{equation}
Then the Chern-Weil isomorphism is explicitly
\begin{equation}
  \BC[\g]^{G}  \simeq  H^{\bullet}(\CP^{\infty}, \BC) \simeq \BC[\ep]
\end{equation}
where $\ep \in \g^{\vee}$ is a linear function on $\g=Lie(S^1)$ and
$\BC[\ep]$ denotes the free polynomial ring on one generator $\ep$.
The  $\ep \in H^2(\CP^{\infty}, \BC)$  is negative of the first Chern class
$c_1$ of the universal bundle 
\begin{equation}
-c_1(\gamma) =   \ep =  \frac{1}{2 \pi \sqrt{-1}} \tr_{\mathbf{1}} F_{A}(\gamma)
\end{equation}
where $\tr_{\mathbf{1}}$ denotes trace of the curvature two-form $F_{A}
= dA + A \wedge A$ in the fundamental complex 1-dimensional
representation in which the Lie algebra of $\g = Lie(S^1)$ is represented by
$\imath \BR$. The  cohomological degree of $\ep$  is
\begin{equation}
  \deg \ep = \deg F_{A}(\gamma) = 2
\end{equation}

Generally, for a compact connected Lie group $G$ we  reduce the Chern-Weil
theory to the maximal torus $T \subset G$ and identify 
\begin{equation}
  \BC[\g]^{G} \simeq \BC[\mathfrak{t}]^{W_{G}}
\end{equation}
where $\mathfrak{t}$ is the Cartan Lie algebra $\mathfrak{t} = Lie(T)$
and $W_{G}$ is the Weyl group of $G$. 

For example, if $G = U(n)$ the Weyl group $W_{U(n)}$ is the permutation
group of $n$ eigenvalues $\ep_1, \dots \ep_n$. Therefore 
\begin{equation}
H^\bullet(BU(n), \BC) =  \BC[\g]^{U(n)} \simeq \BC[\ep_1, \dots, \ep_n]^{W_{U(n)}} \simeq
  \BC[c_1, \dots, c_n]
\end{equation}
where $(c_1, \dots, c_n)$ are elementary symmetrical monomials called
Chern classes 
\begin{equation}
  c_k = (-1)^{k} \sum_{i_1 \leq \dots \leq i_k} \ep_{i_1} \dots
  \ep_{i_k} 
\end{equation}
The classifying space for $G = U(n)$ is 
\begin{equation}
  BU(n) = \lim_{k \to \infty} \mathrm{Gr}_{n}( \BC^{k+n})
\end{equation}
where $\mathrm{Gr}_n(V)$ denotes the space of $n$-planes in the vector space
$V$. 

To summarize, if $G$ is a connected compact Lie group with Lie algebra
$\g = Lie(G)$, maximal torus $T$ and its Lie algebra $\mathfrak{t} =
Lie(T)$, and Weyl group $W_{G}$, then it holds 
\begin{equation}
\boxed{  H_{G}^\bullet(pt, \BR) \simeq H^\bullet(BG, \BR) \simeq \BR[\g]^{G} \simeq \BR[\mathfrak{t}]^{W_{G}}}
\end{equation}

\section{Weil algebra}

The cohomology $H^\bullet(BG, \BR)$ of the classifying space $BG$ can
also be realized in the Weil algebra 
\begin{equation}
 \mathcal{W}_{\g} := \BR[ \g[1] \oplus  \g[2]] = \Lambda\g^{\vee}
 \otimes  S\g^{\vee}
\end{equation}
Here $ \g[1]$ denotes shift of degree so that
elements of $\g[1]$ are Grassmann. The space of polynomial functions 
 $\BR[\g[1]]$  on $\g[1]$ is the anti-symmetric algebra  $ \Lambda\g^{\vee}$ of
 $\g^{\vee}$,  and the space
 of polynomial functions $\BR[\g[2]]$ on $\g[2]$ is the symmetric algebra
 $S \g^{\vee}$  of  $\g^{\vee}$.

The elements $c \in  \g[1]$ have degree 1 and represent the connection
1-form on the universal bundle. The elements $\phi \in \g[2]$ have degree
$2$ and represent the curvature 2-form on the universal bundle.  An odd differential on functions
on $\g[1] \oplus  \g[2]$ can be described as an odd vector field $\delta$
such that $\delta^2=0$.  The odd vector field $\delta$ of degree 1 represents
de Rham differential on the universal bundle
\begin{equation}
\label{PEeq:delta}
  \begin{aligned}
 \delta c &= \phi - \frac 1 2 [c, c] \\
  \delta \phi &= -[c, \phi]  
  \end{aligned}
\end{equation}
which follows from the standard relations between the connection $A$ and
the curvature $F_{A}$ 
\begin{equation}
\label{PEeq:deltaAF}
  \begin{aligned}
  dA &= F_A  - \frac 1 2 [A, A]  \\
  dF_A & = -[A, F_A] 
  \end{aligned}
\end{equation}

This definition implies $\delta^2  = 0$. Indeed,
\begin{equation}
  \begin{aligned}
\delta^2 c & = \delta \phi - [\delta c, c] =-[c,\phi] - [\phi - \frac 1 2
[c,c], c] = 0\\
  \delta^2 \phi &= -[\delta c, \phi] + [c, \delta \phi]  = -[\phi - \frac
  1 2 [c,c], \phi] - [c, [c, \phi] ] =0 
  \end{aligned}
\end{equation}

Given a  basis $T_{\alpha}$ on the Lie algebra $\g$ with structure
constants $[T_\beta, T_{\gamma}]=f^{\alpha}_{\beta\gamma} T_{\alpha}$
the differential $\delta$ has the form
\begin{equation}
  \begin{aligned}
  \delta c^{\alpha} &=  \phi^{\alpha} - \frac 1 2 f^{\alpha}_{\beta
    \gamma} c^{\beta} c^{\gamma}\\
  \delta \phi^{\alpha} &= -f^{\alpha}_{\beta \gamma} c^{\beta}
  \phi^{\gamma}
  \end{aligned}
\end{equation}

The differential $\delta$ can be decomposed into the sum of two
differentials
\begin{equation}
\label{PEeq:combined}  \delta = \delta_{\mathrm{K}} + \delta_{\mathrm{BRST}}
\end{equation}
with
\begin{equation}
  \begin{aligned}
    \delta_{\mathrm{K}} \phi = 0, \qquad \delta_{\mathrm{BRST}} \phi = -[c,\phi] \\
    \delta_{\mathrm{K}} c = \phi, \qquad \delta_{\mathrm{BRST}} c= -\frac 1 2 [c,c]
  \end{aligned}
\end{equation}

The differential $\delta_{\mathrm{BRST}}$ is the BRST differential
(Chevalley-Eilenberg differential for Lie algebra cohomology with
coefficients in the Lie algebra module $S\g^{\vee}$).  The
differential $\delta_{\mathrm{K}}$ is the Koszul differential (de Rham differential
on $\Omega^{\bullet}(\Pi \g)$). 

 The field theory interpretation of the Weil algebra and the 
differential (\ref{PEeq:combined}) was given in \cite{PEKanno:1988wm} and \cite{PELabastida:1988qb}.

The Weil algebra $\mathcal{W}_{\g} = \BR[ \g[1] \oplus \g[2]]$ is
an extension of  the Chevalley-Eilenberg algebra $CE_{\g} = \BR[\g[1]]= \Lambda\g^{\vee}$ by
the algebra $\BR[\g[2]]^{G} = S\g^{\vee}$ of
symmetric  polynomials on $\g$
\begin{equation}
 CE_{\g} \leftarrow \mathcal{W}_{\mathfrak{g}} \leftarrow S\g^{\vee}
\end{equation}
which is quasi-isomorphic to  the algebra of differential forms on the
universal bundle
\begin{equation}
  G \to EG \to BG
\end{equation}
The duality between the Weil algebra $\mathcal{W}_{\mathfrak{g}}$ and the de
Rham algebra $\Omega^{\bullet}(EG)$ of differential forms on $EG$ is
provided by the Weil homomorphism
\begin{equation}
  \mathcal{W}_{\mathfrak{g}} \to \Omega^{\bullet}(EG)
\end{equation}
 after a choice of a connection 1-form $A \in \Omega^{1}(EG) \otimes \g$
 and its field strength $F_{A} \in \Omega^{2}(EG) \otimes \g$ on
the universal bundle $EG \to BG$. 

Indeed,  the connection 1-form $A
\in \Omega^{1}(EG) \otimes \g$ and field strength $F \in
\Omega^{2}(EG)\otimes \g$ define  maps $\g^{\vee} \to \Omega^{1}(EG)$ and
$\g^{\vee} \to \Omega^{2}(EG)$
\begin{equation}
  \begin{aligned}
&   c^{\alpha} \mapsto A^{\alpha} \\
&\phi^{\alpha} \mapsto F^{\alpha}
  \end{aligned}
\end{equation}

The cohomology of the Weil algebra is trivial
\begin{equation}
  H^{n}(\mathcal{W}_{\mathfrak{g}}, \delta, \BR) =
  \begin{cases}
    \BR,\qquad n = 0 \\
    0, \qquad n > 0 
  \end{cases}
\end{equation}
corresponding to the trivial cohomology of $\Omega^{\bullet}(EG)$. 

To define $G$-equivariant cohomology we need to
consider $G$ action on $EG$. 
To compute $H_{G}^{\bullet}(pt) = H^{\bullet}(BG)$, consider 
$\Omega^{\bullet}(BG) = \Omega^{\bullet}(EG/G)$. 

For any principal $G$-bundle $\pi: P \to P/G$ the differential forms on
$P$ in the image of the pullback $\pi^{*}$ of the space of differential forms on $P/G$
are called \emph{basic}
\begin{equation}
  \Omega^{\bullet}(P)_{\mathrm{basic}} = \pi^{*} \Omega^{\bullet}(P/G)
\end{equation}

Let $L_{\alpha}$ be the Lie derivative in the direction of a vector field $\alpha$
generated by a basis element $T_{\alpha} \in \g$, and $i_\alpha$  be the
contraction with the vector field generated by $T_{\alpha}$. 

An element $\omega \in \Omega^{\bullet}(P)_{\mathrm{basic}}$ can be
characterized by two conditions 
\begin{enumerate}
\item $\omega$ is invariant on $P$ with respect to the $G$-action:
  $L_{\alpha} \omega = 0$
\item $\omega$ is horizontal on $P$ with respect to the $G$-action:
  $i_{\alpha} \omega = 0$
\end{enumerate}

In the Weil model the contraction operation $i_{\alpha}$ is realized as 
\begin{equation}
  \begin{aligned}
  i_{\alpha} c^{\beta} = \delta_\alpha^{\beta}\\
  i_{\alpha} \phi^{\beta} = 0
  \end{aligned}
\end{equation}

and the Lie derivative $L_{\alpha}$ is defined by the usual relation
\begin{equation}
  L_{\alpha} = \delta i_{\alpha} + i_{\alpha} \delta
\end{equation}

From the definition of $\Omega^{\bullet}(P)_{\mathrm{basic}}$ for the
case of $P = EG$  we obtain
\begin{equation}
\label{PEeq:HGpoint}
H^\bullet_{G}(pt) =   H^\bullet(BG, \BR) = H^\bullet(\Omega^{\bullet}(EG)_{\mathrm{basic}} , \BR)=
  H^\bullet(\mathcal{W}_{\mathfrak{g}}, \delta, \BR)_{\mathrm{basic}} = (S \g^{\vee})^{G}
\end{equation}

\section{Weil model and Cartan model of equivariant cohomology}

The isomorphism
\begin{equation}
  H(BG, \BR) = H(EG, \BR)_{\mathrm{basic}} =
  H(\mathcal{W}_{\mathfrak{g}}, \delta, \BR)_{\mathrm{basic}}
\end{equation}
suggests to replace the topological model for $G$-equivariant
cohomologies of real manifold  $X$
\begin{equation}
  H_{G}(X, \BR) = H( (X \times EG)/G, \BR)
\end{equation}
by the \emph{Cartan model}
\begin{equation}
    H_{G}(X, \BR) = H( (\Omega^{\bullet}(X) \otimes
S \g^{\vee})^G,  \BR) 
\end{equation}
or by the equivalent algebraic \emph{Weil model}
\begin{equation}
  H_{G}(X, \BR) = H( (\Omega^{\bullet}(X) \otimes
  \mathcal{W}_{\g})_{\mathrm{basic}},  \BR)
\end{equation}

\subsection{Cartan model}
Here $(\Omega^{\bullet}(X) \otimes
S \g^{\vee})^G$ denotes the $G$-invariant subspace in $(\Omega^{\bullet}(X) \otimes
S \g^{\vee})$ under the $G$-action induced from
 $G$-action on $X$ and adjoint  $G$-action on $\g$. 

It is convenient to think about 
$(\Omega^{\bullet}(X) \otimes S \g^{\vee})$ as the space
\begin{equation}
\label{PEeq:Omegacartan}  \Omega^{\bullet, 0}_{C^{\infty}, \mathrm{poly}}(X \times \g)
\end{equation}
of smooth differential forms on $X \times \g$ of degree 0 along
$\g$ and polynomial along $\g$. 

In  $(T_{a})$  basis on $\g$, an element $\phi \in \g$ is represented as
$\phi = \phi^{\alpha} T_{\alpha}$. Then $(\phi^{\alpha})$ is the dual
basis of $\g^{\vee}$. Equivalently $\phi^{\alpha}$ is a linear coordinate on $\g$. 

The commutative ring $\BR[\g]$ of polynomial functions on the vector space
underlying $\g$ is naturally represented in the coordinates as the ring of polynomials in generators
$\{\phi^{\alpha}\}$
\begin{equation}
  \BR[\g] = \BR[\phi^{1}, \dots, \phi^{\mathrm{rk}\, \g} ] 
\end{equation}
Hence, the space (\ref{PEeq:Omegacartan}) can be equivalently presented as
\begin{equation}
  \Omega^{\bullet, 0}_{C^{\infty}, \mathrm{poly}}(X \times \g) =
  \Omega^{\bullet}(X) \otimes \BR[\g]  
\end{equation}

Given an action of the group $G$ on any manifold
$M$
\begin{equation}
\rho_{g}:  m \mapsto g \cdot m
\end{equation}
the induced action on the space of differential forms
$\Omega^{\bullet}(M)$ comes from the pullback by the map $\rho_{g^{-1}}$
\begin{equation}
\label{PEeq:rhog-action}
 \rho_{g}: \omega \mapsto \rho_{g^{-1}}^{*} \omega, \qquad \omega \in \Omega^{\bullet}(M)
\end{equation}
In particular, if $M = \g$ and $\omega \in \g^{\vee}$ is a linear function
on $\g$, then (\ref{PEeq:rhog-action}) is the \emph{co-adjoint action} on $\g^{\vee}$.

The invariant subspace $ ( \Omega^{\bullet}(X) \otimes \BR[\g]
)^G$ forms a  complex with respect to the \emph{Cartan
  differential}
\begin{equation}
\label{PEeq:Cartandiff}
 d_{G} = d \otimes 1 + i_{\alpha} \otimes \phi^{\alpha} 
\end{equation}
where $d: \Omega^{\bullet}(X) \to \Omega^{\bullet +1} (X) $ is the de Rham differential, and
$i_{\alpha}: \Omega^{\bullet}(X) \to \Omega^{\bullet -1 }(X)$   is the
operation of contraction of the vector field on $X$ generated by $T_{\alpha} \in
\g$ with  differential forms in $\Omega^{\bullet}(X)$.

The Cartan model of the $G$-equivariant cohomology  $H_{G}(X)$ is 
\begin{equation}
  H_{G}(X) = H\left( (\Omega^{\bullet}(X) \otimes
  \BR[\g])^G, d_{G}\right)
\end{equation}

To check that $d_{G}^2=0$  on $(\Omega^{\bullet}(X) \otimes
  \BR[\g])^G$ we compute $d_{G}^2$ on $\Omega^{\bullet}(X) \otimes
  \BR[\g]$ and find
\begin{equation}
  d_{G}^2 = L_{\alpha} \otimes \phi^{\alpha}
\end{equation}
where $L_{\alpha}: \Omega^{\bullet}(X) \to \Omega^{\bullet}(X)$ is the
Lie derivative on $X$ 
\begin{equation}
  L_{\alpha} = d i_\alpha + i_\alpha d 
\end{equation}
along vector field generated by $T_{\alpha}$.

 The infinitesimal action by a Lie algebra generator $T_{a}$ on an
element $\omega \in \Omega^{\bullet}(X) \otimes R[\g]$ is 
\begin{equation}
  T_{\alpha} \cdot \omega = (L_\alpha \otimes 1 + 1 \otimes L_{\alpha}) \cdot \omega 
\end{equation}
where $L_{\alpha}\otimes 1 $ is the geometrical Lie derivative by the vector
field generated by $T_\alpha$ on $\Omega^{\bullet}(X)$ and
$1 \otimes L_a$ is the coadjoint action on $\BR[\g]$
\begin{equation}
  L_\alpha  = f_{\alpha \beta}^{\gamma} \phi^{\beta}
  \frac{\partial}{\partial \phi^{\gamma}}
\end{equation}

If $\omega$ is a $G$-invariant element, $\omega \in (\Omega^{\bullet}(X)
\otimes R[\g])^{G}$,  then
\begin{equation}
\label{PEeq:Lie-geom-alg}
 (L_{\alpha} \otimes 1 + 1 \otimes L_{\alpha}) \omega  = 0
\end{equation}

Therefore, if $\
 \omega \in (\Omega^{\bullet}(X)
\otimes R[\g])^{G}$ it holds
that\begin{equation}
  d_{G}^2 \omega =  (1 \otimes \phi^{\alpha}L_\alpha)\omega = \phi^{\alpha} f^{\gamma}_{\alpha\beta}
  \phi^{\beta} \frac {\partial \alpha}{\partial \phi^{c}} = 0
\end{equation}
by the antisymmetry of the structure constants  $f^{\gamma}_{\alpha\beta} =
-f^{\gamma}_{\beta\alpha}$. Therefore $d_{G}^2 = 0$  on 
 $(\Omega^{\bullet}(X) \otimes   \BR[\g])^G$.

The grading on $\Omega^\bullet(X) \otimes \BR[\g]$ is defined  by the assignment 
\begin{equation}
\label{PEeq:grading}
  \deg d = 1 \quad \deg i_{v_\alpha} = -1 \quad \deg \phi^{\alpha} = 2
\end{equation}
which implies 
\begin{equation}
  \deg d_{G} = 1
\end{equation}
Let 
\begin{equation}
\Omega^{n}_{G}(X) = \oplus_{k} (\Omega^{n-2k} \otimes \BR[\g]^{k})^{G}
  \end{equation}
be the subspace in $ (\Omega(X) \otimes \BR[\g])^{G}$ of degree $n$
according to the grading (\ref{PEeq:grading}).

Then 
\begin{equation}
\label{PEeq:Cartan-complex}
\dots \stackrel{d_G}{\to}  \Omega^{n}_G (X) \stackrel{d_G}{\to}
\Omega^{n+1}_G(X) \stackrel{d_G}{\to} \dots   
\end{equation}
is a differential complex. The equivariant cohomology groups
$H_{G}^{\bullet}(X)$  in the Cartan model are
defined as the cohomology of the complex~(\ref{PEeq:Cartan-complex})
\begin{equation}
  H_{G}^{\bullet}(X) \equiv \operatorname{Ker} d_{G} / \operatorname{Im} d_{G}
\end{equation}

In particular, if $X =pt $ is a point then 
\begin{equation}
  H^{\bullet}_{G}(pt) = \BR[\g]^{G}
\end{equation}
in agreement with (\ref{PEeq:HGpoint}). 

If $x^{\mu}$ are coordinates on $X$, and $\psi^{\mu} = d x^{\mu}$ are
Grassman coordinates on the fibers of $\Pi TX$, we can represent the
Cartan differential (\ref{PEeq:Cartandiff}) in the notations more common in quantum field theory
traditions 
\begin{equation}
  \begin{aligned}
  \delta  x^{\mu} = \psi ^{\mu} \\
 \delta  \psi^{\mu} = \phi^{\alpha} v_{\alpha}^{\mu}
  \end{aligned}\qquad   \delta \phi = 0
\label{eq:PECartandiffcoor}
\end{equation}
where $v^{\mu}$ are components of the vector field on $X$  generated by a basis
element $T_{\alpha}$ for the $G$-action on $X$. In quantum field theory,
the coordinates $x^{\mu}$ are typically coordinates on the
infinite-dimensional space of bosonic fields, and $\psi^{\mu}$ are
typically coordinates on the infinite-dimensional space of fermionic
fields. 

\subsection{Weil model}
The differential in Weil model can be presented in coordinate notations
similar to (\ref{eq:PECartandiffcoor}) as follows
\begin{equation}
\begin{aligned}
  \delta x^{\mu} &= \psi ^{\mu} +c^{\alpha} v_\alpha^{\mu} \\
  \delta \psi^{\mu} &= \phi^{\alpha} v_{\alpha}^{\mu} +
 \partial_{\nu} v^{\mu}_\alpha   c^{\alpha} \psi^{\nu}
\end{aligned}
\qquad 
\begin{aligned}
  \delta c^{\alpha} &=  \phi^{\alpha} - \frac 1 2 f^{\alpha}_{\beta
    \gamma} c^{\beta} c^{\gamma}\\
  \delta \phi^{\alpha} &= -f^{\alpha}_{\beta \gamma} c^{\beta}
  \phi^{\gamma}
\end{aligned}
  \end{equation}

In physical applications, typicallly $c$ is the BRST ghost field for
gauge symmetry, and Weil differential is the sum of a supersymmetry
transformation and BRST transformation, for example see \cite{PEPestun:2007rz}.

\section{Equivariant characteristic classes in Cartan model}
\label{PEse:equivcharclass}
For a reference see \cite{PEBott:2001} and
\cite{PEMR1288997}. 

Let $G$ and $T$ be compact connected Lie groups.

 We consider a $T$-equivariant $G$-principal bundle $\pi: P \to
X$. This means that an equivariant $T$-action is defined on $P$ compatible with the
$G$-bundle structure of $\pi: P \to X$. One can take that $G$ acts from
the right and $T$ acts from the left. 

The compatibility means that $T$-action on the total space of $P$
\begin{itemize}
\item commutes with the projection map $\pi: P \to X$
\item commutes with the $G$ action on the fibers of $\pi: P \to X$
\end{itemize}

Let $D_A = d + A$ be a $T$-invariant connection on a $T$-equivariant
$G$-bundle $P$. Here the connection $A$ is a $\g$-valued $1$-form on
the total space of $P$ (such a connection always exists by the averaging procedure for compact $T$).

 Then we define the $T$-equivariant
connection 
\begin{equation}
   D_{A,T} = D_{A} + \ep^{a} i_{v_a} 
\end{equation}
and the $T$-equivariant curvature
\begin{equation}
  F_{A,T} = (D_{A,T})^2 - \ep^{a} \otimes \mathcal{L}_{v_a} 
\end{equation}
where $\ep^{a}$ are coordinates on the Lie algebra $\mathfrak{t}$
(like  the coordinates $\phi^{a}$ on the Lie algebra
$\mathfrak{g}$ in the previous section defining Cartan model of
$G$-equivariant cohomology), 
which is in fact is an element of $\Omega_{T}^{2}(X) \otimes \g$
\begin{equation}
\label{PEeq:equivariant-curvature}
  F_{A,T} = F_{A}  -  \ep^{a} \otimes \mathcal{L}_{v_a} + [ \ep^{a} \otimes i_{v_a}, 1
  \otimes D_{A}] = F_{A} + \ep^{a} i_{v_a} A 
\end{equation}

  Let $X^{T}$ be the $T$-fixed point set in $X$.
If the equivariant curvature $F_{A,T}$ is evaluated on $X^{T}$,  only the vertical component of $i_{v_a}$
contributes to the formula (\ref{PEeq:equivariant-curvature}) and $v_a$
pairs with the vertical component of the connection
$A$ on the $T$-fiber of $P$ given by $g^{-1} d g$. The $T$-action on
$G$-fibers induces the homomorphism 
\begin{equation}
\label{PEeq:rhog}  \rho: \mathfrak{t} \to \mathfrak{g}
\end{equation}
and let  $\rho(T_a)$ be the images of $T_a$ basis elements of $\mathfrak{t}$.

An ordinary characteristic class for a principal $G$-bundle on $X$ is
$[p(F_A)] \in H^{2d}(X)$ for a $G$-invariant degree $d$ polynomial
$p \in \BR[\g]^{G}$. Here $F_A$ is the curvature of any connection $A$ on
the $G$-bundle. 

In the same way, a $T$-equivariant characteristic class for a principal
$G$-bundle associated to a $G$-invariant degree $d$ polynomial $p
\in \BR[\g]^G$ is $[p(F_{A,T})] \in H_{T}^{2d}(X)$. Here  $F_{A,T}$ is
the $T$-equivariant curvature of any $T$-equivariant connection $A$ on
the $G$-bundle. 

Restricted to $T$-fixed points $X^{T}$ the $T$-equivariant
characteristic class associated to polynomial $p \in \BR[\g]^{G}$ is 
\begin{equation}
\label{PEeq:equivclass}
  p(F_{A} + \ep^{a} \rho(T_a)) 
\end{equation}

In particular, if  $V$ is a representation of $G$ and $p$ is the Chern character of the vector
bundle $V$, then if $X$ is a point, the equivariant Chern characters is
an ordinary character of the space $V$ as a $G$-module. 

\section{Standard characteristic classes}

For a reference see the book by Bott and Tu \cite{PEBott:book}.

\subsection{Euler class}
  Let $G=SO(2n)$ be the special orthogonal group which preserves a Riemannian
  metric $g \in S^2 V^{\vee}$
  on an oriented real vector space $V$ of $\dim_{\BR} V = 2n $.

The Euler
  characterstic class is defined by the adjoint invariant polynomial
  \begin{equation}
    \Pf: \mathfrak{so}(2n, \BR) \to \BR
  \end{equation}
 of degree $n$
  on the Lie algebra $\mathfrak{so}(2n)$ called \emph{Pfaffian} and
  defined as follows. For an element $x \in \mathfrak{so}(2n)$ let $x'
  \in V^{\vee} \otimes V$  denote representation of $x$ on $V$
  (fundamental representation), so that $x'$ is an antisymmetric $(2n)\times(2n)$ matrix in some orthonormal basis of $V$.  Let $g \cdot x'
  \in \Lambda^2 V^{\vee}$ be the two-form associated by $g$ to $x'$ , and let
  $v_{g} \in \Lambda^{2n} V^{\vee}$ be the standard volume form on $V$
  associated to the metric $g$, and $v^*_{g} \in \Lambda^{2n} V$ be the dual
  of $v_{g}$. 
By definition 
\begin{equation}
\label{PEeq:Pf} \Pf(x)  = \frac{1}{n!}\langle v_{g}^{*}, (g \cdot x')^{\wedge n} \rangle
\end{equation}

For example, for the $2 \times 2$-blocks diagonal matrix 
\begin{equation}
  \Pf
  \begin{pmatrix}
    0 & \ep_1 & \dots & \dots & 0 & 0   \\
   -\ep_1 &  0  & \dots & \dots & 0 & 0 \\
   \dots   & \dots & \dots & \dots & \dots & \dots \\
   \dots   & \dots & \dots & \dots & \dots & \dots \\
   0      &  0 & \dots & \dots & 0 & \ep_n \\
   0 &  0 &  \dots & \dots & -\ep_n & 0 
  \end{pmatrix}
 = \ep_1 \dots \ep_n
\end{equation}

For an antisymmetric $(2n) \times (2n)$ matrix $x'$, the definition
implies that  $\Pf(x)$ is a  degree $n$ polynomial of matrix elements of $x$ which satisfies  
\begin{equation}
  \Pf(x)^2 = \det x
\end{equation}

Let $P$ be an $SO(2n)$ principal bundle $P \to X$. 

In the standard normalization the Euler class $\euler(P)$ is defined in such
a way that it takes values in $H^{2n}(X, \BZ)$ and is given by 
\begin{equation}
\label{PEeq:Euler1}
  \euler(P) = \frac{1}{(2\pi)^n} [\Pf( F)]
\end{equation}

For example, the Euler characteristic of an oriented real manifold $X$ of real dimension
$2n$ is an integer number given by
\begin{equation}
\label{PEeq:Euler}
  \euler(X) = \int_{X} \euler(T_{X})  = \frac{1}{ (2 \pi)^{n}} \int_{X} \Pf (R)
\end{equation}
where $R$ denotes the curvature form of the tangent bundle $T_{X}$. 

In quantum field theories the definition  (\ref{PEeq:Pf}) of the Pfaffian is
usually realized in terms of a Gaussian
integral over the Grassmann (anticommuting) variables $\theta$ which 
satisfy $\theta_i \theta_j = -\theta_j \theta_i$. The definition
(\ref{PEeq:Pf}) is presented as 
\begin{equation}
\Pf(x) =  \int d \theta_{2n} \dots d \theta_{1} \exp( - \frac 1 2 \theta_{i}
  x_{ij} \theta_j) 
\end{equation}
By definition, the integral
$[d\theta_{2n} \dots d\theta_{1}]$ picks the coefficient of the monomial
$\theta_1 \dots \theta_{2n}$ of an element of the 
the Grassman algebra generated by $\theta$.

\subsection{Euler class of vector bundle and Mathai-Quillen form}
See Mathai-Quillen \cite{PEMathaiQuillen} and Aityah-Jeffrey
\cite{PEMR1094734}. 

The Euler class of a vector bundle can be presented in a QFT formalism. 
Let $E$ be an oriented real vector bundle $E$ of rank $2n$ over a manifold $X$.  

Let
$x^{\mu}$ be local coordinates on the base $X$, and let their differentials
be denoted $\psi^{\mu} = d x^{\mu}$. 

Let $h^{i}$ be local
coordinates on the fibers of $E$. Let $\Pi E$ denote the superspace
obtained from the total space of the bundle $E$ by inverting the parity
of the fibers, so that the coordinates in the fibers of $\Pi E$ are odd
variables $\chi^{i}$.   Let $g_{ij}$ be the matrix of a Riemannian metric on the bundle
$E$.  Let $A^{i}_{\,\mu}$ be the matrix valued 1-form on $X$
representing a connection on the bundle $E$. 

Using the connection $A$ we can define an odd vector field $\delta$ on the superspace
$\Pi T ( \Pi E)$, or, equivalently, a de Rham differential on the space of
differential forms $\Omega^{\bullet} (\Pi E)$. In local coordinates
$(x^{\mu}, \psi^{\mu})$ and $(\chi^{i}, h^{i})$ the definition of
$\delta$ is
\begin{equation}
  \begin{aligned}
  \delta x^{\mu} &= \psi^{\mu} \\
  \delta \psi^{\mu} &= 0 
  \end{aligned} \quad 
  \begin{aligned}
    \delta \chi^{i} &= h^{i} - A^{i}_{\,j\mu} \psi^{\mu} \chi^{j} \\
    \delta h^{i} &= \delta (A^{i}_{\,j\mu} \psi^{\mu} \chi^{j})
  \end{aligned}
\end{equation}

Here $h^{i} = D \chi^{i}$ is the \emph{covariant} de Rham differential
of $\chi^{i}$, so that under the change of framing on $E$ given by
$\chi^{i} = s^{i}_{\,j} \tilde \chi^{j}$ the $h^{i}$ transforms
in the same way, that is $h^{i} = s^{i}_{\, j} \tilde h^{j}$. 

The odd vector field $\delta$ is nilpotent
\begin{equation}
  \delta^2 = 0
\end{equation}
and is called de Rham vector field on $\Pi T(\Pi E)$. 

Consider an element $\alpha$ of  $\Omega^{\bullet}(\Pi E)$ defined by
the equation 
\begin{equation}
\label{PEeq:alphadef}
  \alpha = \frac{1}{(2 \pi)^{2n}} \exp(  - t   \delta V)
\end{equation}
where $t \in \BR_{> 0}$ and 
\begin{equation}
  V = \frac 1 2 (g_{ij} \chi^{i} h^{j})
\end{equation}
Notice that since $h^{i}$ has been defined as $D\chi^i$ the definition
(\ref{PEeq:alphadef}) is coordinate independent. 

To expand the definition of $\alpha$ (\ref{PEeq:alphadef}) we compute
\begin{equation}
  \delta (\chi, h) = (h - A\chi, h) - (\chi, dA \chi - A (h - A\chi))=
  (h,h) - (\chi, F_{A} \chi)
\end{equation}
where we suppresed the indices $i,j$, the $d$ denotes the de Rham
differential on $X$ and $F_{A}$ the curvature $2$-form on the connection
$A$
\begin{equation}
  F_{A} = d A + A \wedge A
\end{equation}

The Gaussian integration of the form $\alpha$ along the vertical fibers of $\Pi E$ gives
\begin{equation}
\label{PEeq:MQ}
 \frac{1}{(2\pi)^{2n}} \int [dh][d\chi]\exp (-\frac 1 2 \delta (\chi, h))
 = \frac{1}{(2 \pi)^n} \Pf (F_{A})
\end{equation}
which agrees with definition of the integer valued Euler class
(\ref{PEeq:Euler1}). The representation of the Euler class in the form (\ref{PEeq:alphadef}) is called
the Gaussian \emph{Mathai-Quillen representation} of the Thom class. 

The Euler class of the vector bundle $E$ is an element of $H^{2n}(X,
\BZ)$. If $\dim X = 2n$, the number obtained after integration of the
fundamental cycle on $X$ 
\begin{equation}
\label{PEeq:Enumber}
  e(E)  = \int_{\Pi T (\Pi E)}  \alpha
\end{equation}
is an integer Euler characterstic of the vector bundle $E$.

If $E = TX$ the equation (\ref{PEeq:Enumber}) provides the Euler
characteristic of the manifold $X$ in the form 
\begin{equation}
  e(X)  = \frac{1}{(2\pi)^{\dim X}} \int_{\Pi T (\Pi TX)} \exp(-t  \delta V
  )  \stackrel{ t \to 0} {=}  \frac{1}{(2\pi)^{\dim X}}   \int_{\Pi T (\Pi
    TX)} 1 
\end{equation}

Given  a  section $s$ of the vector bundle $E$,  we can deform the form
$\alpha$ in the same $\delta$-cohomology class by taking
\begin{equation}
\label{PEeq:Vsection}
   V_{s}=   \frac  1 2 (\chi, h + \sqrt{-1} s) 
\end{equation}

After integrating over $(h, \chi)$ the the resulting differential form
on $X$ has factor 
\begin{equation}
  \exp( -\frac{1}{2 t} s^2)
\end{equation}
so it is concentraited in a neigborhood of the locus $s^{-1}(0) \subset X$
of  zeroes of the section $s$. 

In this way the Poincare-Hopf theorem is proven: given an oriented vector bundle $E$ on an oriented manifold $X$, with $\rank E
= \dim X$, the Euler characteristic of $E$ is equal to the number of
zeroes of a generic section $s$ of $E$ counted with orientation
\begin{equation}
  e(E) = \sum_{x \in s^{-1}(0) \subset X}  \mathrm{sign} \det ds|_{x}
\end{equation}
where $ds|_{x}: T_{x}  \to E_{x} $ is the differential of the section $s$ at a zero $x
\in s^{-1}(0)$. The assumption that $s$ is a generic section implies that
$\det ds|_{x}$ is non-zero.

For a short reference on the Mathai-Quillen formalism see \cite{PEWu:2005pr}. 

\subsection{Chern character}

Let $P$ be a principal $GL(n,\BC)$ bundle over a manifold $X$. The Chern
character is an adjoint invariant function 
\begin{equation}
 \ch: \mathfrak{gl}(n,\BC)
\to \BC 
\end{equation}
 defined as the trace in the fundamental representation of the
exponential map 
\begin{equation}
\ch:   x \mapsto \tr e^{x}
\end{equation}
The exponential map is defined by formal series
\begin{equation}
 \tr  e^{x} = \sum_{n=0}^{\infty}\frac{1}{n!} \tr x^{n}
\end{equation}

The eigenvalues of the $gl(n,\BC)$  matrix $x$ are called \emph{Chern
  roots}. In terms of the Chern roots the Chern character is 
\begin{equation}
 \ch(x) = \sum_{i=1}^{n}  e^{x_i} 
\end{equation}

\subsection{Chern class}
Let $P$ be a principal $GL(n,\BC)$ bundle over a manifold $X$.
The Chern class $c_k$ for $k
\in \BZ_{>0}$ of $x \in \mathfrak{gl}(n,\BC)$ is defined by expansion of the determinant 
\begin{equation}
 \det (1 + t x)  = \sum_{k=0}^{n}  t^{n} c_n
\end{equation}
In particular
\begin{equation}
  c_1(x) = \tr x, \qquad c_n(x) = \det x 
\end{equation}

In terms of Chern roots $c_k$ is defined by elementary symmetric
monomials 
\begin{equation}
  c_k = \sum_{ 1 \leq  i_1 < i_2 \dots < i_k \leq n} x_{i_1} \dots x_{i_n}
\end{equation}

\emph{Remark on integrality.}  Our conventions for characteristic classes of $GL(n,\BC)$ bundles 
differ from the frequently used conventions in which Chern classes
$c_k$ take value in $H^{2k}(X, \BZ)$ by a factor of $(-2 \pi
\sqrt{-1})^{k}$. 
 In our conventions the  characteristic class of degree $2k$ needs to be
 multiplied by $\frac{1}{
(-2 \pi \sqrt{-1})^{k}}$ to be integral.

\subsection{Todd class}
Let $P$ be a principal $GL(n,\BC)$ bundle over a manifold $X$.
The Todd class of $x \in \mathfrak{gl}(n,\BC)$ is defined to be 
\begin{equation}
  \td (x) = \det  \frac{x}{1 - e^{-x}}   = \prod_{i=1}^n \frac{ x_i}{ 1- e^{-x_i}}
\end{equation}
where $\det$ is evaluated in the fundamental representation. 
The ratio evaluates to a series expansion involving Bernoulli
numbers $B_k$
\begin{equation}
  \frac{x}{1-e^{-x}} = \sum_{k=0}^{\infty} \frac{(-1)^{k}}{k!} B_k x^{k}
  = 1 + \frac{x}{2} + \frac{x^2}{12} - \frac{x^4}{720} + \dots 
\end{equation}

\subsection{The $\hat A$ class}
Let $P$ be a principal $GL(n,\BC)$ bundle over a manifold $X$.
The $\hat A$ class of $x \in GL(n, \BC)$ is defined as 
\begin{equation}
 \hat A  = \det \frac{x}{e^{\frac x 2} - e^{-\frac x 2}}=   \prod_{i=1}^{n} \frac{ x_i}{ e^{x_i /2} - e^{-x_i/2}}
\end{equation}
The $\hat A$ class is related to the Todd class by
\begin{equation}
\label{PEeq:ATd}
   \hat A (x) = \det e^{-\frac x 2} \td x  
\end{equation}

\section{Index formula}
For a holomorphic vector bundle $E$ over a complex variety $X$ of
$\dim_{\BC} X = n$ the index $\ind(\bar \partial, E)$ is defined as
\begin{equation}
\label{PEeq:inde}
\ind(\bar \partial, E) =
\sum_{k=0}^{n} (-1)^k \dim H^{k}(X,
E) 
\end{equation}

The localization theorem in $K$-theory 
gives the index formula of Grothendieck-Riemann-Roch-Hirzebruch-Atiyah-Singer 
relating the index to the Todd class
\begin{equation}
\label{PEeq:Toddindex}
\boxed{
  \ind (\bar \partial, E) = \frac{1}{(-2 \pi \sqrt{-1})^n} \int_{X} \td (T_X^{1,0}) \ch(E) }
\end{equation}

Similarly, the index of Dirac operator $\slashed{D}: S^{+}\otimes E \to
S^{-}\otimes E$  from the positive chiral
spinors $S^{+}$ to the negative chiral  spinors $S^{-}$,  twisted by a
vector bundle $E$, is defined as 
\begin{equation}
\ind(\slashed{D}, E) = \dim \ker \slashed{D} - \dim \coker \slashed{D} 
\end{equation}
and is given by the Atiyah-Singer index formula 
\begin{equation}
\label{PEeq:equivariant-Dirac}
\boxed{  \ind (\slashed{D}, E)  = \frac{1}{(-2 \pi \sqrt{-1})^n} \int_{X} \hat A (T_X^{1,0}) \ch(E) }
\end{equation}

Notice that on a Kahler manifold the Dirac complex $$\slashed{D}: S^{+} \to S^{-}$$
is isomorphic  to the Dolbeault
complex  $$\dots \to \Omega^{0,p}(X)
\stackrel{\bar \partial}{\to} \Omega^{0,p+1}(X) \to \dots $$ twisted by
the square root of the canonical bundle $K = \Lambda^{n} (T_X^{1,0})^{\vee}$
\begin{equation}
\label{PEeq:DDolb}
  \slashed{D}  = \bar \partial \otimes K^{\tfrac 1 2}
\end{equation}
consistently with the relation  (\ref{PEeq:ATd}) 
and the Riemann-Roch-Hirzebruch-Atiyah-Singer index formula

\emph{Remark on $2\pi$ and $\sqrt{-1}$ factors.}
The vector bundle $E$ in the index formula (\ref{PEeq:Toddindex}) can be
promoted to a complex 
\begin{equation}
\to E^{\bullet} \to E^{\bullet+1} \to 
\end{equation}

In particular, the $\bar \partial$ index of the complex $E^\bullet = \Lambda^\bullet
(T^{1,0})^{\vee}$ of $(\bullet,0)$-forms on a Kahler variety $X$ equals
the Euler characteristic of $X$
\begin{equation}
e(X)  =   \ind(\bar \partial,  \Lambda^\bullet (T^{1,0})^{\vee}) =  \sum_{q=0}^{n}
  \sum_{p=0}^n (-1)^{p+q} \dim H^{p,q}(X) 
\end{equation}
We find 
\begin{equation}
  \ch  \Lambda^\bullet (T^{1,0})^{\vee} = \prod_{i=1}^{n} (1 - e^{-x_i})
\end{equation}
where $x_i$ are Chern roots of the curvature of the $n$-dimensional
complex bundle $T_{X}^{1,0}$. Hence, the Todd index formula
(\ref{PEeq:Toddindex}) gives
\begin{equation}
\label{PEeq:ec-agreement}
  \euler(X) = \frac{1}{(2 \pi \sqrt{-1})^{n}} \int c_n(T_{X}^{1,0})
\end{equation}

The above agrees with  the Euler characteristic (\ref{PEeq:Euler}) provided it holds
that 
\begin{equation}
\label{PEeq:Pf-det}
 \det ( \sqrt{-1} x_{\mathfrak{u}(n)}) =   \Pf(x_{\mathfrak{so}(2n)}) 
\end{equation}
where $x_{\mathfrak{so}(2n)}$ represents the curvature of the $2n$-dimensional
real tangent bundle $T_{X}$ as $2n \times 2n$ antisymmetric matrices, 
and $x_{\mathfrak{u}(n)}$ represents the curvature of the complex
holomorphic $n$-dimensional tangent bundle $T_{X}^{(1,0)}$ as $n \times
n$ anti-hermitian matrices. 
That (\ref{PEeq:Pf-det}) holds is clear from the $(2\times 2$ representation of
$\sqrt{-1}$
\begin{equation}
  \sqrt{-1} \mapsto
  \begin{pmatrix}
    0 & -1 \\
    1 & 0
  \end{pmatrix}
\end{equation}

\section{Equivariant integration} 

See the paper by Atiyah and Bott \cite{PEAtiyah1984}. 

\subsection{Thom isomorphism and Atiyah-Bott localization}

A map $$f: F \to X$$ of  manifolds induces a natural
 pushfoward map on the homology $$ f_{*} : H_{\bullet}(F) \to
 H_{\bullet}(X)$$ and pullback on the cohomology $$ f^{*}:
 H^{\bullet}(X) \to H^{\bullet}(F)$$

In the situation when there is Poincar\'e duality between homology and
cohomology we can construct pushforward operation on the cohomology
\begin{equation}
  f_{*} : H^{\bullet}(F) \to H^{\bullet}(X)
\end{equation}
We can display the pullback and pushforward maps on the diagram
\begin{equation}
  H^{\bullet}(F) \substack{\overset{f_{*}}{\longrightarrow} \\ \underset{f^{*}}{\longleftarrow}} H^{\bullet}(X)
\end{equation}
For example, if $F$ and $X$ are compact manifolds and $f: F \hookrightarrow X$ is
the inclusion, then for the pushforward map $f_{*}: H^{\bullet}(F)  \to
H^{\bullet}(X)$  we find
\begin{equation}
\label{PEeq:f1}
  f_{*} 1 = \Phi_{F}
\end{equation}
where $\Phi_{F}$  is the cohomology class in $H^{\bullet}(X)$ which is Poincar\'e
dual to the manifold $F \subset  X$:  for a form $\alpha$ on $X$
we have
\begin{equation}
\label{PEeq:Poincare}
  \int_{F} f^{*} \alpha = \int_{X} \Phi_{F} \wedge \alpha 
\end{equation}
If $X$ is the total space of the orthogonal vector bundle $\pi: X \to F$
over the oriented manifold
$F$ then $\Phi_F(X)$ is called the Thom class of the vector bundle $X$ and $f_{*}: H^{\bullet}(F)
\to H^{\bullet}(X)$ is the Thom isomorphism: to a form $\alpha$ on $F$
we associate a form $\Phi \wedge \pi^{*} \alpha$ on $X$. 
The important property of the Thom class $\Phi_F$ for a submanifold $F \hookrightarrow X$ is
\begin{equation}
  f^{*} \Phi_F = \euler(\nu_F)
\end{equation}
where $\euler(\nu_F)$ is the Euler class of the normal bundle to $F$ in
$X$. Combined with (\ref{PEeq:f1}) the last equation gives
\begin{equation}
  f^{*} f_{*} 1 = \euler(\nu_F)
\end{equation}
as a map $H^{\bullet}(F) \to H^{\bullet}(F)$.

More generally, if $f: F \hookrightarrow X$ is an inclusion of a manifold $F$
into a manifold $X$ the Poincar\'e dual class $\Phi_{F}$ is isomorphic to the
Thom class of the normal bundle of $F$ in $X$. 

Now we consider $T$-equivariant cohomologies for a compact abelian Lie group $T$
acting on $X$. Let  $F =X^{T}$ be the set of  $T$ fixed points in
$X$. Then the equivariant Euler class $\euler_{T}(\nu_F)$ is invertible, 
therefore the identity map on $H^{\bullet}_T(X)$ can be presented as 
\begin{equation}
1 = f_{*} \frac{1}{\euler_T(\nu_F)}  f^{*} 
\end{equation}
Let $\pi^{X}: X \to pt$ be the map from a manifold $X$ to a point $pt$. The
pushforward operator $\pi^{X}_{*}: H^{\bullet}_T(X) \to H^{\bullet}_T(pt)$
corresponds to the integration of the cohomology class over
$X$. The pushforward is functorial. For maps $F \stackrel{f}{\to} X
\stackrel{\pi^X}{\to} pt$ we have the composition $\pi^X_* f_* = \pi^{F}_*$ 
for $F \stackrel{\pi^{F}}{\to} pt$. So we arrive to the Atiyah-Bott integration formula
\begin{equation}
  \pi^X_* = \pi^F_*  \frac{f^*}{\euler_T(\nu_F)}
\end{equation}
or more explicitly 
\begin{equation}
\label{PEeq:ABlocalization}
\boxed{
  \int_{X} \alpha = \int_{F} \frac{ f^{*} \alpha}{\mathrm{\euler}_{T}(\nu_F)}
}
\end{equation}

\subsection{Duistermaat-Heckman localization}

A particular example where the Atiyah-Bott localization formula can be applied
is a symplectic space on which a Lie group $T$ acts in a Hamiltonian
way. Namely, let $(X, \omega)$ be a real symplectic manifold of
$\dim_{\BR} X = 2n$ 
 with
symplectic form $\omega$ and let compact connected Lie group $T$ act on $X$ in Hamiltonian
way, which means that there exists a function, called \emph{moment map} or Hamiltonian
\begin{equation}
  \mu: X \to \mathfrak{t}^{\vee}
\end{equation}
such that 
\begin{equation}
  d \mu_a = -i_{a} \omega
\end{equation}
in some basis $(T_{a})$ of $\mathfrak{t}$ where $i_{a}$ is the contraction
operation with the vector field generated by the $T_{a}$ action on $X$. 

The degree 2 element $\omega_{T} \in \Omega^\bullet(X) \otimes S\mathfrak{t}^{*}$ defined by
the equation
\begin{equation}
 \omega_{T} =  \omega + \ep^{a} \mu_a 
\end{equation}
is  a $d_{T}$-closed equivariant differential form:
\begin{equation}
d_{T} \omega_{T} = (  d + \ep^{a} i_{a} ) (\omega + \ep^{b} \mu_b) = \ep^{a} d\mu_{a} +
\ep^{a} i_{a} \omega  = 0
\end{equation}

This implies that the mixed-degree equivariant differential form 
\begin{equation}
  \alpha = e^{\omega_{T}}
\end{equation}
is also $d_{T}$-closed, and we can apply the Atiyah-Bott localization
formula to the integral 
\begin{equation}
\label{PEeq:DH-integral}
  \int_{X} \exp(\omega_{T}) = \frac{1} {n!} \int_{X} \omega^{n}
 \exp(\ep^{a} \mu_a)
\end{equation}

For $T = SO(2)$ so that $\mathrm{Lie}(SO(2)) \simeq \BR$ the integral (\ref{PEeq:DH-integral}) is the typical partition function of
a classical Hamiltonian mechanical system in statistical physics with
Hamiltonian function $\mu: X \to \BR$ and inverse temperature parameter
$-\ep$. 

Suppose that  $T = SO(2)$ and that the set of fixed points $X^{T}$ is
discrete. Then the Atiyah-Bott localization formula
(\ref{PEeq:ABlocalization}) implies
\begin{equation}
\label{PEeq:DH}
\boxed{
\frac{1} {n!} \int_{X} \omega^{n}
 \exp(\ep^{a} \mu_a) =  \sum_{x \in X^{T}} \frac{\exp(\ep^{a} \mu_a)}{\euler_{T}(\nu_x)}}
\end{equation}
where $\nu_x$ is the normal bundle to a fixed point $x \in X^{T}$ in
$X$ and $\euler_{T}(\nu_x)$ is  the $T$-equivariant Euler class of the bundle
$\nu_x$.   

The rank of the normal bundle $\nu_x$ is $2n$ and the
structure group is $SO(2n)$. In notations of section \ref{PEse:equivcharclass} we
evaluate the $T$-equivariant characteristic Euler class of the principal
$G$-bundle for $T = SO(2)$ and $G=SO(2n)$ by equation
(\ref{PEeq:equivclass}) for the invariant polynomial  on $\mathfrak{g} =
\mathfrak{so}(2n)$ given by $p = \frac{1}{(2\pi)^{n}} \mathrm{Pf}$
according to definition (\ref{PEeq:Euler1}).

\subsection{Gaussian integral example}
To illustrate the localization formula (\ref{PEeq:DH})  suppose that $X = \BR^{2n}$ with symplectic form
\begin{equation}
  \omega = \sum_{i=1}^{n} dx^{i} \wedge dy_i
\end{equation}
and $SO(2)$ action 
\begin{equation}
\label{PEeq:SO2action}
  \begin{pmatrix}
    x_i \\
    y_i 
  \end{pmatrix} \mapsto 
  \begin{pmatrix}
    \cos w_i \theta & -\sin w_i\theta \\
    \sin w_i \theta & \cos w_i \theta 
  \end{pmatrix}
  \begin{pmatrix}
    x_i \\
    y_i 
  \end{pmatrix}
\end{equation}
where $\theta \in
\BR/(2\pi \BZ)$ parametrizes $SO(2)$ and $(w_1, \dots, w_n) \in
\BZ^{n}$. 

The point $0 \in X$ is the \emph{fixed point} so that $X^{T} = \{0\}$,
and the normal bundle $\nu_{x} = T_{0} X$ is an $SO(2)$-module of real
dimension $2n$ and complex dimension $n$ that splits into a direct sum 
of $n$ irreducible $SO(2)$ modules  with weights $(w_1, \dots,
w_n)$.  

We identify $Lie(SO(2))$ with $ \BR$ with basis element $\{1\}$ and
coordinate function $\ep \in Lie(SO(2))^{*} $. The $SO(2)$
action (\ref{PEeq:SO2action}) is Hamiltonian with respect to the moment
map 
\begin{equation}
  \mu =\mu_0 + \frac{1}{2} \sum_{i=1}^{n} w_i (x_i^2 + y_i^2)
\end{equation}

Assuming that $\ep < 0$ and all $w_i > 0$ we find by direct Gaussian integration 
\begin{equation}
  \frac{1}{n!} \int_{X} \omega^{n} \exp( \ep \mu) =
  \frac{(2\pi)^{n}}{(-\ep)^n \prod_{i=1}^{n} w_i} \exp(  \ep \mu_0)
\end{equation}
and the same result  by the localization formula (\ref{PEeq:DH}) because
\begin{equation}
  \euler_{T}(\nu_{x}) = \frac{1}{(2\pi)^n}\Pf (  \ep \rho(1))
\end{equation}
according to the definition of the $T$-equivariant class (\ref{PEeq:equivclass}) 
and the Euler characteristic class (\ref{PEeq:Euler1}), and  
where $\rho: Lie(SO(2)) \to Lie(SO(2n))$ is the homomorphism in
(\ref{PEeq:rhog}) with 
\begin{equation}
  \rho(1) =
  \begin{pmatrix}
    0 & -w_1 & \dots & \dots & 0 & 0   \\
   w_1 &  0  & \dots & \dots & 0 & 0 \\
   \dots   & \dots & \dots & \dots & \dots & \dots \\
   \dots   & \dots & \dots & \dots & \dots & \dots \\
   0      &  0 & \dots & \dots & 0 & -w_n \\
   0 &  0 &  \dots & \dots & w_n & 0     
  \end{pmatrix}
\end{equation}
according to (\ref{PEeq:SO2action}).

\subsection{Example of a two-sphere} 

Let $(X,\omega)$ be the two-sphere $S^2$ with coordinates $(\theta,
\alpha)$ and symplectic structure
\begin{equation}
  \omega = \sin \theta  d \theta \wedge d \alpha
\end{equation}
Let the Hamiltonian function be 
\begin{equation}
  H = - \cos \theta  
\end{equation}
so that
\begin{equation}
  \omega = dH \wedge d \alpha
\end{equation}
and the Hamiltonian vector field be $v_{H} = \partial_\alpha$. The
differential form 
$$\omega_T  =  \omega + \ep H = \sin \theta d \theta \wedge d \alpha
-\ep \cos \theta $$ is $d_T$-closed for 
\begin{equation}
  d_{T} = d + \ep i_{\alpha}
\end{equation}
Let 
\begin{equation}
  \alpha = e^{ t \omega_{T}}
\end{equation}
 Locally there is a degree $1$ form $V$ such that  $\omega_T
= d_T V$, for example
\begin{equation}
   V = - (\cos \theta) d \alpha
\end{equation}
but globally $V$  does not exist. The $d_T$-cohomology class
$[\alpha]$ of the form $\alpha$ is non-zero. 

The localization formula (\ref{PEeq:DH-integral}) gives
\begin{equation}
  \int_{X} \exp( \omega_{T}) = \frac{2\pi}{-\ep}\exp(-\ep) +
  \frac{2\pi}{\ep} \exp(\ep)
\end{equation}
where the first term is the contribution of the  $T$-fixed point
$\theta = 0$ and the second term is the contribution of the $T$-fixed
point $\theta = \pi$.

\section{Equivariant index formula (Dolbeault and Dirac)}
Let $G$ be a compact connected Lie group.  

Suppose that $X$ is a complex variety and $E$ is a holomorphic $G$-equivariant vector
bundle over $X$. Then the cohomology groups $H^\bullet(X, E)$ form
representation of $G$.   In this case the index of  $E$ (\ref{PEeq:inde}) can be refined to 
\emph{an equivariant index} or \emph{character}
\begin{equation}
\label{PEeq:equivindex}
\ind_{G}(\bar \partial, E) =
\sum_{k=0}^{n} (-1)^k \ch_{G} H^{k}(X,
E) 
\end{equation}
where $\ch_{G} H^{i}(X,E)$ is the character of a representation of $G$ in
the vector space $H^{i}(X, E)$. More concretely, the equivariant index can
be thought of as a gadget that attaches to $G$-equivariant holomorphic
bundle $E$ a complex valued adjoint invariant function on the group
$G$
\begin{equation}
\label{PEeq:equivindex2}
  \ind_{G}(\bar \partial, E)(g) = \sum_{k=0}^{n} (-1)^k \tr_{H^{k}(X,E)} g
\end{equation}
on elements $g \in G$. The sign alternating sum (\ref{PEeq:equivindex2})
is also known as 
\emph{the supertrace}
\begin{equation}
    \ind_{G}(\bar \partial, E)(g) = \str_{H^{\bullet}(X,E)} g
\end{equation}

The index formula (\ref{PEeq:Toddindex}) is replaced by the equivariant index
formula in which  characteristic classes are promoted to 
$G$-equivariant characteristic classes in the Cartan model of
$G$-equivariant cohomology with differential $d_{G} = d + \phi^{a} 
i_{a}$ as in (\ref{PEeq:Cartandiff}) 
\begin{equation}
\label{PEeq:equivindex3}
   \ind (\bar \partial, E)(e^{\phi^a T_a})  = 
\frac{1} {(-2 \pi \sqrt{-1})^n} \int_{X} \td_{G} (T_X) \ch_{G}(E)  =
\int_{X} e_{G}(T_X) \frac{ \ch_{G} E}{\ch_{G} \Lambda^{\bullet} T_{X}^{\vee}}
 \end{equation}
Here $\phi^{a} T_{a}$ is an element of Lie algebra of $G$ and
$e^{\phi^a T_{a} }$ is an element of $G$, and $T_{X}$ denotes the
holomorphic tangent bundle of the complex manifold $X$.

If the set $X^{G}$ of $G$-fixed points is discrete, then applying  the localization formula (\ref{PEeq:ABlocalization}) to the equivariant
index (\ref{PEeq:equivindex3})  we find the equivariant Lefshetz formula
\begin{equation}
\label{PEeq:Dolindex}
\boxed{     \ind (\bar \partial, E)(g )  = \sum_{ x \in X^{G}} \frac{ \tr_{E_{x}}(g) }{ \det_{T^{1,0}_{x}X}(1 - g^{-1})}}
\end{equation}
The Euler character is cancelled against the numerator of the Todd
character.

\subsubsection{Example of $\CP^1$} Let $X$ be $\CP^{1}$ and let $E
=\mathcal{O}(n)$ be a complex
line bundle of degree $n$ over $\CP^{1}$, and let $G = U(1)$
equivariantly act on $E$ as follows. 
Let $z$ be a local coordinate on $\CP^1$, and let an element $t \in U(1)
\subset \BC^{\times}$ send the point with coordinate $z$ to the point with
coordinate $t z$ so that 
\begin{equation}
  \ch T_{0}^{1,0} X  = t \qquad   \ch T_{\infty}^{1,0} X  = t^{-1}
\end{equation}
where $T^{1,0}_0 X$ denotes the fiber of the holomorphic tangent bundle at
$z=0$ and similarly $T^{1,0}_\infty X$  the fiber at $z=\infty$.
 Let the action of $U(1)$ on the fiber of $E$ at $z=0$ be
trivial. Then the action of $U(1)$ on the fiber of $E$ at $z =\infty$ is
found from the gluing relation 
\begin{equation}
  s_\infty = z^{-n} s_0
\end{equation}
to be of weight $-n$, so that 
\begin{equation}
  \ch E|_{z=0} = 1, \qquad \ch E|_{z=\infty}  = t^{- n}
\end{equation}

Then 
\begin{equation}
\label{PEeq:chfibers}
  \ind(\bar \partial, \mathcal{O}(n), \CP^1)(t) =  \frac{1}{1 - t^{-1}}
  + \frac{t^{-n}}{1 - t} = \frac{1 - t^{-n-1}}{1 - t^{-1}} =   \begin{cases}
    \sum_{k=0}^{n} t^{-k}, \quad n \geq 0 \\
    0, \quad n = -1,  \\
    -t\sum_{k=0}^{-n-2} t^{k}, \quad n < -1
  \end{cases}
\end{equation}

We can check against the direct computation. Assume $n \geq 0$. 
The kernel of $\bar \partial$ is spanned by $n+1$ holomorphic sections of
$\mathcal{O}(n)$ of the form $z^k$ for $k = 0, \dots, n$, the cokernel is empty by
Riemann-Roch. 
The section $z^k$ is acted upon by $t \in T$ with weight $t^{-k}$. Therefore
\begin{equation}
\label{PEeq:CP1example}
  \ind_T(\bar \partial , \mathcal{O}(n), \CP^{1}) = \sum_{k=0}^{n} t^{-k} 
\end{equation}

Even more explicitly, for illustration, choose
a connection 1-form  $A$ with constant curvature $F_A = - \frac 1 2 i n
\omega$, denoted in the patch around $\theta = 0$ (or $z=0$) by $A^{(0)}$
and in the patch around $\theta = \pi$ (or $z=\infty$) by $A^{(\pi)}$
\begin{equation}
    A^{(0)} = -\frac 1 2 i n ( 1 - \cos \theta) d \alpha \qquad     A^{(\pi)}
    = -\frac 1 2 i n ( - 1 - \cos \theta) d \alpha \\
\end{equation}
The gauge transformation between the two patches 
\begin{equation}
  A^{(0)} = A^{(\pi)} - i n\, d \alpha 
\end{equation}
is consistent with the defining $E$ bundle transformation
rule for the sections $s^{(0)}, s^{(\pi)}$ in the patches around $
\theta = 0$ and $\theta = \pi$ 
\begin{equation}
  s^{(0)} = z^n s^{(\pi)} \qquad A^{(0)} =  A^{(\pi)} + z^{n} d z^{-n} .
\end{equation}

The equivariant curvature $F_T$ of the connection $A$ in the 
bundle $E$ is given by
\begin{equation}
\label{PEeq:Finv}
  F_{T} = -\frac 1 2 i n (\omega + \ep (1- \cos \theta))
\end{equation}
as can be verified against the definition
(\ref{PEeq:equivariant-curvature}) $F_{T} =  F + \ep i_{v} A$.
Notice that to verify the expression for the equivariant
curvature~(\ref{PEeq:Finv}) in the patch near $\theta  = \pi$ one needs to take into
account contributions from the vertical component $g^{-1} dg$ of the connection $A$ on the total space
of the principal $U(1)$ bundle and from the $T$-action on the
fiber at $\theta = \pi $ with weight $-n$.

Then
\begin{equation}
  \begin{aligned}
\ch (E)|_{\theta =0}  &=   \exp (F_{T})|_{\theta = 0} = 1 \\
\ch (E)|_{\theta = \pi}  &=  \exp (F_{T})|_{\theta = \pi} = \exp(-in
\ep)  = t^{-n}
  \end{aligned}
\end{equation}
for $t = \exp( i \ep)$ in agreement with (\ref{PEeq:chfibers}). 

A similar exercise gives the index for the Dirac operator on $S^2$ twisted by
a magnetic field of flux $n$
\begin{equation}
\label{PEeq:Dindex}
  \ind(\slashed{D}, \mathcal{O}(n), S^2) = \frac{t^{n/2} -
    t^{-n/2}}{t^{\frac 1 2} - t^{-\frac 1 2}}
\end{equation}
where now we have chosen the lift of the $T$-action symmetrically to be
of weight $n/2$ at $\theta =0$ and of weight $-n/2$ at $\theta
=\pi$. Also notice that up to overall multiplication by a power of $t$
related to the choice of lift of the $T$-action to the fibers of the bundle $E$, the
relation (\ref{PEeq:DDolb}) holds
\begin{equation}
\label{PEeq:shift}
  \ind(\slashed{D}, \mathcal{O}(n), S^2)  =   \ind(\bar \partial, \mathcal{O}(n-1), \CP^1)
\end{equation}
because on $\CP^1$ the canonical bundle is $K =\mathcal{O}(-2)$.

\subsubsection{Example of $\CP^m$}.   Let $X = \CP^{m}$ be defined by the projective coordinates $(x_0:x_1:\dots
  :x_m)$ and $L_n$ be the line bundle $L_n =
  \mathcal{O}(n)$. Let  $T =U(1)^{(m+1)}$ act
  on  $X$ by
  \begin{equation}
    (x_0:x_1:\dots x_m) \mapsto (t_0^{-1} x_0: t_1^{-1} x_1: \dots: t_m^{-1} x_m)
  \end{equation}
and by $t_{k}^{n}$ on the fiber of the bundle $L_n$ in the patch 
around the $k$-th fixed point $x_{k}=1, x_{i \neq k} = 0$.  We find the
index as a sum of contributions from $m+1$ fixed points
\begin{equation}
\label{PEeq:fixed}
  \ind_{T}(D) = \sum_{k=0}^{m} \frac{ t_k^{n}}{ \prod_{j\neq k} (1 - (t_j/t_k))}
\end{equation}
For $n \geq 0$ the index is a homogeneous polynomial in
$\BC[t_0, \dots, t_m]$  of degree $n$ representing the character on the space of
holomorphic sections of the $\mathcal{O}(n)$ bundle over $\CP^m$. 
\begin{equation}
\label{PEeq:cp-answer}
  \ind_{T}(D) =
  \begin{cases}
    s_{n}(t_0, \dots, t_m), \quad n \geq 0 \\
    0, \quad   -m  \leq   n  < 0 \\
    (-1)^m t_0^{-1} t_1^{-1} \dots t_m^{-1}  s_{-n-m-1}(t_0^{-1}, \dots, t_m^{-1}), \quad   n
    \leq -m - 1
  \end{cases}
\end{equation}
where $s_n(t_0,\dots, t_m)$ are complete homogeneous symmetric
polynomials. 
This result can be quickly obtained from the contour integral
representation of the sum (\ref{PEeq:fixed})
\begin{equation}
  \frac{1}{2 \pi i} \oint_{\mathcal{C}} \frac{dz}{z} \frac{ z^{n}}{ \prod_{j=0}^{m} (1
    - t_j / z)} = 
  \sum_{k=0}^{m} \frac{ t_k^{n}}{ \prod_{j\neq k} (1 - (t_j/t_k))} ,
\end{equation}
If $ n
\geq -m $ we pick the contour of integration $\mathcal{C}$ to  enclose all residues
$z=t_j$. The residue at $z = 0$ is zero and the sum of
residues is (\ref{PEeq:fixed}). On the other hand, the same contour
integral is evaluated by the residue at $z = \infty$ which is computed
by expanding all fractions in inverse powers of $z$, and is given by 
the complete homogeneous polynomial in $t_i$ of degree $n$.

If $n < -m$ we assume that the contour of integration is a small circle
around the $z=0$ and does not include any of the residues $z =
t_j$. Summing the residues outside of the contour, and taking that $z=\infty$ does not contribute, we get 
(\ref{PEeq:fixed}) with the  ($-$) sign . The residue at
$z = 0$ contributes by (\ref{PEeq:cp-answer}).

Also notice that the last line of (\ref{PEeq:cp-answer})
relates\footnote{Thanks to Bruno Le Floch for the comment} to the
first line by the reflection $t_i \to t_i^{-1}$
\begin{equation}
  \frac{t_k^n}{\prod_{j\neq k}(1-t_j/t_k)} = \frac{(-1)^m (t_k^{-1})^{-n-m-1} (\prod_j t_j^{-1}) }{ \prod_{j\neq k} (1-t_j^{-1}/t_k^{-1})}
\end{equation}
which is the consequence of the Serre duality on $\CP^{m}$. 

\section{Equivariant index and representation theory}
The $\CP^1$ in  example (\ref{PEeq:Dindex})  can be thought of as a flag manifold $SU(2)/U(1)$, and
(\ref{PEeq:chfibers}) (\ref{PEeq:Dindex}) as characters of $SU(2)$-modules. 
For index theory on general flag manifolds $G_{\BC}/B_{\BC}$, 
that is Borel-Weyl-Bott theorem\footnote{For a short presentation see
  exposition by J. Lurie at  \url{http://www.math.harvard.edu/~lurie/papers/bwb.pdf}}, the shift of the form 
(\ref{PEeq:shift}) is a shift by the Weyl vector $ \rho = \sum_{\alpha > 0}
\alpha$ where $\alpha$ are positive roots of $\mathfrak{g}$. 

The index formula with localization to the fixed points on a flag manifold
is equivalent to the Weyl character formula.

The generalization of formula (\ref{PEeq:Dindex}) for generic flag
manifold appearing from a co-adjoint orbit in $\mathfrak{g}^{*}$ is called
\emph{Kirillov character formula} \cite{PEMR2069175}, \cite{PEMR0294568},
\cite{PEMR1701415}. 

Let $G$ be a compact simple Lie group. 
The Kirillov character formula equates  the $T$-equivariant index of the Dirac operator $\ind_T(D)$ on the $G$-coadjoint
orbit of the element $\lambda + \rho
\in \g^{*}$ with the character $\chi_{\lambda}$  of the $G$ irreducible representation with
highest weight $\lambda$.

The character $\chi_{\lambda}$ is a function $\g \to \BC$ determined by the
representation of the Lie group $G$ with highest weight $\lambda$
as
\begin{equation}
  \chi_{\lambda}:  X \mapsto \tr_\lambda e^{X}, \qquad X \in \g
\end{equation}
Let  $X_{\lambda}$ be an orbit of the
co-adjoint action by $G$ on $\g^{*}$.  Such orbit is specified by an
element $\lambda \in \mathfrak{t}^{*}/W$ where $\mathfrak{t}$ is the Lie
algebra of the maximal torus $T \subset G$  and $W$
is the Weyl group. The co-adjoint orbit $X_\lambda$ is a homogeneous symplectic
$G$-manifold with the canonical symplectic structure $\omega$ defined at point
$x \in X \subset \g^{*}$ on tangent vectors in $\g$ by the formula
\begin{equation}
 \omega_x(\bullet_1, \bullet_2) =   \la x, [ \bullet_1, \bullet_2 ] \ra \qquad \bullet_1, \bullet_2 \in \g
\end{equation}
The converse is also true: any homogeneous symplectic
$G$-manifold is locally isomorphic to a coadjoint orbit of $G$ or
central extension of it.

The minimal possible stabilizer of $\lambda$ is the maximal abelian
subgroup $T \subset G$, and the
 maximal co-adjoint orbit is $G/T$. Such orbit is called \emph{a full flag
manifold}. The real dimension of the full flag manifold is  $2n = \dim G
- \rk G$, and is  equal to  the number of  roots of $\g$.  If the
stabilizer of  $\lambda$ is a larger group $H$, such that $T \subset H
\subset G$, 
the orbit $X_{\lambda}$ is called \emph{a partial flag manifold}
$G/H$.
 A degenerate flag manifold is a projection from the full flag manifold with fibers isomorphic to $H/T$. 

Flag manifolds are equipped with natural complex and Kahler structure. 
 There is an explicitly holomorphic  realization of the
flag manifolds as a complex quotient $G_{\BC}/P_{\BC}$ where $G_{\BC}$ is
the complexification of the compact group $G$ and $P_{\BC} \subset
G_{\BC}$ is a  parabolic
subgroup.
Let  $\g = \mathfrak{g}_{-}
\oplus \mathfrak{h} \oplus \mathfrak{g}_{+}$ be the standard decomposition of $\mathfrak{g}$ into the Cartan
$\mathfrak{h}$ algebra and the upper triangular $\g_{+}$ and lower
triangular $\g_{-}$  subspaces. 

 The minimal parabolic subgroup is known as Borel subgroup
$B_{\BC}$, its  Lie algebra is conjugate to  $\mathfrak{h} \oplus
\mathfrak{g_{+}}$.  The Lie algebra of  generic parabolic subgroup $P_{\mathbb{C}} \supset
B_{\mathbb{C}}$ is conjugate
to the direct sum of $\mathfrak{h} \oplus \mathfrak{g}_{+}$  and a proper subspace of $\mathfrak{g}_{-}$. 

Full flag manifolds with integral symplectic structure are in
bijection with irreducible $G$-representations $\pi_{\lambda}$ of
highest weight $\lambda$
\begin{equation}
  X_{\lambda + \rho} \leftrightarrow \pi_{\lambda}
\end{equation}
This is known as the Kirillov correspondence in geometric representation theory.

Namely, if $\lambda \in \g^{*}$ is a weight,  the symplectic
 structure $\omega$ is integral and there exists a line bundle $L \to
 X_{\lambda}$ with a unitary connection of curvature $\omega$.  The
 line bundle $L \to X_{\lambda}$ is acted upon by the maximal torus $T
 \subset G$ and we can study the
$T$-equivariant geometric objects. 
The Kirillov-Berline-Getzler-Vergne character formula 
equates the equivariant index of the Dirac
operator $\slashed{D}$ twisted by the line bundle $L \to X_{\lambda + \rho}$ on the
co-adjoint orbit $X_{\lambda + \rho}$
with the character $\chi_{\lambda}$ of the irreducible representation of $G$
with  highest weight $\lambda$ 
\begin{equation}
\ind_{T}(\slashed{D})(X_{\lambda + \rho}) =  \chi_{\lambda} 
\end{equation}
This formula can be easily proven using the Atiyah-Singer equivariant
index formula 
 \begin{equation}
\ind_{T}(\slashed{D})(X_{\lambda + \rho}) = 
\frac{1}{(-2 \pi i)^{n}}  \int_{X_{\lambda+\rho}}  \ch_{T}(L) \hat A_{T}(T_{X})
 \end{equation}
and the Atiyah-Bott formula to localize the integral over $X_{\lambda +
  \rho}$ to the set of fixed points $X^{T}_{\lambda + \rho}$.

The localization to  $X^{T}_{\lambda+\rho}$
yields the Weyl formula for the character.
Indeed, the stabilizer of $\lambda + \rho$, where $\lambda$ is a
dominant weight, is the Cartan torus $T \subset G$. The co-adjoint orbit 
$X_{\lambda + \rho}$ is the full flag manifold. The $T$-fixed
points are in the intersection  $X_{\lambda + \rho} \cap \mathfrak{t}$, and hence, the set of the $T$-fixed
points is the Weyl orbit of $\lambda + \rho$
\begin{equation}
  X^{T}_{\lambda + \rho} = \mathrm{Weyl}(\lambda + \rho)
\end{equation}

At each fixed point $p \in X^{T}_{\lambda + \rho}$  the tangent space $T_{X_{\lambda+\rho}}|_{p}$ is
generated by the root system of $\mathfrak{g}$. The tangent space is a
complex $T$-module $\oplus_{\alpha > 0} \BC_{\alpha}$ with weights
$\alpha$ given by  the positive roots of $\mathfrak{g}$. Consequently, the denominator of  $\hat A_{T}$ gives the
Weyl denominator, the numerator of  $\hat A_{T}$ cancels with the Euler
class $e_{T}(T_{X})$  in the localization formula, and the restriction of 
$\ch_{T}(L) = e^{\omega}$ is  $e^{ w(\lambda+\rho)}$
\begin{equation}
\frac{1}{(-2 \pi i)^n}  \int_{X_{\lambda+\rho}}  \ch_{T}(L)  \hat A(T_{X})
 = \sum_{w \in W} \frac{  e^{i w(\lambda+\rho) \ep}} { \prod_{\alpha > 0} ( e^{\frac 1 2 i \alpha \ep}
     - e^{- \frac 1 2 i \alpha \ep})}
\end{equation}
We conclude  that the localization of the equivariant index of the Dirac
operator on $X_{\lambda+\rho}$ twisted by the line bundle $L$  to the
set of  fixed points $X^{T}_{\lambda + \rho}$ is
precisely the Weyl formula for the character.

The Kirillov correspondence between the index of the Dirac operator of
$L \to X_{\lambda + \rho}$ and the character is closedly related to the 
Borel-Weyl-Bott theorem. 

    Let $B_{\BC}$ be a Borel subgroup of $G_{\BC}$, $T_{\BC}$ be the maximal torus,
$\lambda$ an integral weight of $T_{\BC}$.  A weight $\lambda$ defines a
one-dimensional representation  of $B_{\BC}$ by pulling back the
representation on $T_{\BC} = B_{\BC}/U_{\BC}$ where $U_{\BC}$ is the
unipotent radical of $B_{\BC}$ (the unipotent radical $U_{\BC}$
  is generated by $\mathfrak{g}_{+}$).  Let  $L_{\lambda} \to G_{\BC}/B_{\BC}$ be the associated
line bundle, and $\mathcal{O}(L_{\lambda})$ be the
sheaf of regular local sections of $L_{\lambda}$. 
For $w \in \mathrm{Weyl}_{G}$  define  the action of $w$  on a  weight
$\lambda$ by $w * \lambda := w(\lambda + \rho) - \rho$. 

The \emph{Borel-Weyl-Bott} theorem is that for any weight $\lambda$ one has
\begin{equation}
  H^{l(w)}(G_{\BC}/B_{\BC},\mathcal{O}(L_{\lambda}))
  =
  \begin{cases}
     R_{\lambda}, \quad  w *\lambda \quad \text{is dominant}\\
     0, \quad w *\lambda \quad \text{is not dominant}
  \end{cases}
\end{equation}
where $R_{\lambda}$ is the irreducible $G$-module with  highest
weight $\lambda$, the $w$ is an element of Weyl group such that $w *
\lambda$ is dominant weight, and $l(w)$ is the length of $w$. 
We remark that if there exists $w \in \mathrm{Weyl}_{G}$  such that $w * \lambda$ is
dominant weight, then $w$ is unique.  There is no  $w \in
\mathrm{Weyl}_{G}$ such that $w * \lambda$ is dominant 
if in the basis of the fundamental weights $\Lambda_i$ some of the
coordinates of $\lambda + \rho$ vanish.

\subsubsection{Example}

For $G = SU(2)$ the $G_{\BC}/B_{\BC}  = \CP^1$, an integral weight of $T_{\BC}$ is an
integer $n \in \BZ$, and the line bundle $L_n$ is the $\mathcal{O}(n)$
bundle over $\CP^{1}$.   The Weyl weight is $ \rho = 1$. 

The weight $n \geq 0$  is dominant and the $H^{0}(\CP^{1}, \mathcal{O}(n))$ is the
$\SL(2,\BC)$ module of highest weight $n$ (in the basis of fundamental
weights of $SL(2)$).

For weight $n  = -1$ the $H^{i}(\CP^1, \mathcal{O}(-1))$ is empty for all $i$ as
there is no Weyl transformation $w$ such that $w*n$ is dominant
(equivalently,  because $ \rho + n = 0$). 

For weight $n \leq  -2$ the $w$ is the $\BZ_2$ reflection and $w * n = -(n + 1) - 1
= -n - 2$ is dominant and $H^{1}(\CP^1, \mathcal{O}(n))$ is an
irreducible $SL(2,\BC)$ module of highest weight $ - n - 2$.

The relation between Borel-Weil-Bott theorem for $G_{\BC}/B_{\BC}$ and the Dirac
complex on $G_{\BC}/B_{\BC}$ is that Dirac operator is precisely the Dolbeault
operator shifted by the square root of the canonical bundle
\begin{equation}
  S^{+}(X) \ominus S^{-}(X) = K^{\frac 1 2} \sum (-1)^{p} \Omega^{0,p}(X)
\end{equation}
and consequently 
\begin{equation}
  \ind (X_{\lambda + \rho}, \slashed{D} \otimes L_{\lambda  + \rho})  = \ind
  (G_{\BC}/B_{\BC}, \bar \partial \otimes L_{\lambda})
\end{equation}

The Borel-Bott-Weyl theorem has a generalization for partial flag
manifolds.   Let $P_{\BC}$ be a parabolic subgroup of $G_{\BC}$ with $B_{\BC} \subset P_{\BC}$ and let
$\pi: G_{\BC}/B_{\BC} \to G_{\BC}/P_{\BC}$ denote the canonical
projection. Let $E \to  G_{\BC}/P_{\BC}$ be
a vector bundle associated to an irreducible finite dimensional $P_{\BC}$ module, and
let $\mathcal{O}(E)$ the
the sheaf of local regular sections of $E$. Then
$\mathcal{O}(E)$ is isomorphic to the  direct image
sheaf $\pi_{*} \mathcal{O}(L)$ for a one-dimensional
$B_{\BC}$-module $L$  and
\[
H^{k}(G_{\BC}/P_{\BC}, \mathcal{O}(E)) = H^{k}(G_{\BC}/B_{\BC}, \mathcal{O}(L))
\] 

For application of Kirillov theory to Kac-Moody and Virasoro algebra see
\cite{PEAlekseev:1988ce}.

\section{Equivariant index for differential operators}
See the book by Atiyah \cite{PEAtiyah1974}.

Let $E_k$ be vector bundles over a manifold $X$. Let  $G$ be a
compact Lie group acting on $X$ and the bundles $E_k$.  The action
of $G$ on a bundle $E$ induces canonically a linear action on the space of
sections $\Gamma(E)$. For $g \in G$ and a section $\phi \in \Gamma(E)$ the
action is 
\begin{equation}
\label{PEeq:section-action}
   (g \phi)(x) = g \phi( g^{-1} x ), \quad x \in X
\end{equation}
Let $D_k$ be  linear differential operators compatible with the $G$
action, and let $\mathcal{E}$ be the complex (that is $D_{k+1} \circ D_k
= 0$)
\begin{equation}
\label{PEeq:complex}
\mathcal{E}:  \Gamma(E_0) \stackrel{D_0}{\to} \Gamma(E_1)
\stackrel{D_1}{\to} \Gamma(E_2) \to \dots 
\end{equation}
Since $D_k$ are $G$-equivariant operators, the $G$-action on
$\Gamma(E_k)$ induces the $G$-action on the cohomology $H^{k}(\mathcal{E})$. 
The equivariant index of the complex  $\mathcal{E}$ is the virtual character 
\begin{equation}
  \ind_{G}(D): \mathfrak{g}  \to \BC 
\end{equation}
defined by 
\begin{equation}
  \ind_G(D) (g)=  \sum_{k} (-1)^k \tr_{
    H^{k}(\mathcal{E})}g
\end{equation}

\subsection{Atiyah-Singer equivariant index formula for elliptic complexes}
If the set $X^{G}$ of $G$-fixed points is discrete, the Atiyah-Singer equivariant index formula is
\begin{equation}
\label{PEeq:ASindex}
\boxed{  \ind_{G}(D) = \sum_{x \in X^{G}} \frac{ \sum_{k} (-1)^k
    \ch_G(E_k)|_x }{ \det_{T_{x}X }(1 -g^{-1})}}
\end{equation}

For the Dolbeault complex  $E_k = \Omega^{0,k}$ and $D_k = \bar \partial:
\Omega^{0,k} \to \Omega^{0, k +1}$ 
\begin{equation}
 \to \Omega^{0,\bullet}
\stackrel{\bar \partial}{\to } \Omega^{0, \bullet+1} \to  
\end{equation}
the index (\ref{PEeq:ASindex}) agrees with (\ref{PEeq:Dolindex}) because the
numerator in (\ref{PEeq:ASindex}) decomposes as $\ch_{G} E  \ch_{G}
\Lambda^\bullet T_{0,1}^{*}$ and the denominator as $\ch_{G} \Lambda^\bullet
T_{0,1}^* \ch_{G} \Lambda^{\bullet} T_{1,0}^{*}$ and the factor $\ch_G
\Lambda^{\bullet} T_{0,1}^{*} $ cancels out.

For example,  the equivariant index of $\bar \partial: \Omega^{0,0}(X) \to \Omega^{0,1}(X)$ on $X = \BC_{\la x \ra}$ under the
 $T=U(1)$  action $x \mapsto t^{-1} x$
where $ t \in T$ is the fundamental character is contributed by the
fixed point $x =0$ as 
\begin{equation}
  \ind_{T}(\BC, \bar \partial) = \frac{1 - \bar t}{(1 - t)(1 - \bar
    t)} = \frac{1}{ 1 - t} = \sum_{k = 0}^{\infty} t^{k}
\end{equation}
where the denominator is the determinant of the operator $1 - t$ over the two-dimensional normal
bundle to $0 \in \BC$ spanned by the vectors $\partial_{x}$ and
$\partial_{\bar x}$ with eigenvalues $t$ and $\bar t$. In the numerator,
$1$ comes from the equivariant Chern character on the fiber of
the trivial line bundle at $x=0$ and $-\bar t$ comes from the
equivariant Chern character on the fiber of the bundle of $(0,1)$ forms
$d \bar x$. 

We can compare the expansion in power series in $t^k$  of the index with
the direct computation. The terms $t^{k}$ for $k  \in
\BZ_{\geq 0}$ come from the local $T$-equivariant holomorphic functions
$x^k$ which span the kernel of $\bar \partial$ on $\BC_{\la x \ra}$. The
cokernel is empty by the Poincar\'e lemma. 
Compare with (\ref{PEeq:CP1example}).

Similarly, for the $\bar \partial$ complex on $\BC^{r}$ we obtain 
\begin{equation}
    \ind_{T}(\BC^{r}, \bar \partial) =  \left[\prod_{k=1}^{r} \frac{1}{ (1 - t_k)}\right]_{+}
\end{equation}
where $[]_{+}$ means expansion in positive powers of $t_k$. 

For application to the localization computation on spheres of even
dimension $S^{2r}$ we can compute the index of a certain transversally
elliptic operator $D$ which naturally interpolates between the $ \bar \partial $-complex
in the neighborhood of one fixed point (north pole) of the $r$-torus $T^{r}$ action on
$S^{2r}$ and the $\bar \partial$-complex in the neighborhood of another fixed
point (south pole). The index is a sum of two fixed point
contributions

\begin{equation}
  \begin{aligned}
    \ind_{T}(S^{2r}, D) =  \left[\prod_{k=1}^{r} \frac{1}{
        (1 - t_k)}\right]_{+}
+ \left[\prod_{k=1}^{r} \frac{1}{ (1 - t_k)}\right]_{-}  \\
=  \left[\prod_{k=1}^{r} \frac{1}{
        (1 - t_k)}\right]_{+}
+ \left[\prod_{k=1}^{r} \frac{(-1)^{r} t_1^{-1} \dots t_r^{-1}}{ (1 - t_k^{-1})}\right]_{-}
  \end{aligned}
\end{equation}
where $[]_{+}$ and $[]_{-}$ denotes the expansions in positive and
negative powers of $t_k$.

\subsection{Atiyah-Singer index formula for a free action $G$-manifold}

Suppose that a compact Lie group $G$ acts freely on a manifold $X$ and let $Y =X/G$ be the
quotient, and let 
\begin{equation}
\label{PEeq:Gbundle}
  \pi: X \to Y
\end{equation}
be the associated $G$-principal bundle.

Suppose that $D$ is a $G \times T$ equivariant operator (differential) for a complex 
$(\mathcal{E},D)$ of vector bundles $E_{k}$ over $X$ as in
(\ref{PEeq:complex}). The $G\times T$-equivariance means that the complex
$\mathcal{E}$ and the operator $D$ are pullbacks by $\pi^{*}$ of
a $T$-equivariant complex $\tilde{\mathcal{E}}$ and operator $\tilde D$ on
the base $Y$
\begin{equation}
  \mathcal{E} = \pi^{*} \tilde{\mathcal{E}}, \quad D = \pi^{*} \tilde D 
\end{equation}
We want to compute the $G\times T$-equivariant index   $\ind_{G \times T}(D; X)$ for the complex $(\mathcal{E},D)$
on the total space $X$ for a $G \times T$ transversally elliptic operator $D$ using
$T$-equivariant index theory on the base $Y$. We can do that using Fourier theory on $G$ (counting KK modes
in $G$-fibers).

Let $R_{G}$ be the set of all irreducible representations of $G$. For
each irreducible representation $\alpha \in R_{G}$ we denote
by $\chi_{\alpha}$ the character of this representation, and by $W_{\alpha}$
the vector bundle over $Y$ associated to the principal $G$-bundle
(\ref{PEeq:Gbundle}). 
Then, for each irrep $\alpha \in R_{G}$ we consider a complex $\tilde{\mathcal{E}}\otimes W_{\alpha}$ on $Y$
 obtained by tensoring
$\tilde{\mathcal{E}}$ with the vector bundle $W_{\alpha}$ over $Y$. 
The Atiyah-Singer formula is 
\begin{equation}
\label{PEeq:free-action}
  \ind_{G \times T}(D; X) = \sum_{\alpha \in R_{G}} \ind_{T}(\tilde D
  \otimes W_{\alpha} ;   Y) \chi_{\alpha}.
\end{equation}

\subsubsection{Example of $S^{2r-1}$}
We consider an example immediately relevant for localization on
odd-dimensional spheres $S^{2r-1}$ which are subject to the equivariant
action of the maximal torus $T^{r}$ of the isometry group $SO(2r)$. The
sphere $\pi: S^{2r-1} \to \CP^{r-1}$ is the total space of the $S^1$ Hopf fibration over
the complex projective space $\CP^{r-1}$. 

We will apply the equation (\ref{PEeq:free-action}) for a transversally
elliptic operator $D$ induced from the Dolbeault operator $\tilde D = \bar \partial$
on $\CP^{r-1}$ by the pullback $\pi^{*}$.

To compute the index of operator $D = \pi^{*} \bar \partial$ on
$\pi: S^{2r-1} \to  \CP^{r-1}$ we apply
(\ref{PEeq:free-action}) and use (\ref{PEeq:cp-answer}) and obtain
\begin{equation}
  \ind(D, S^{2r-1}) = \sum_{n=-\infty}^{\infty}
  \ind_{T}(\bar \partial , \CP^{r-1}, \mathcal{O}(n))  =
  \left[\frac{1}{\prod_{k=1}^{r}(1 - t_k)}\right]_{+} + \left[\frac{ (-1)^{r-1} t_1^{-1} \dots t_r^{-1}}{\prod_{k=1}^{r}(1 - t_k^{-1})}\right]_{-}
\end{equation}
where $[]_{+}$ and $[]_{-}$ denotes the expansion in positive and
negative powers of $t_k$.  See further review in \volcite{PZ}.

\subsection{General Atiyah-Singer index formula}

The Atiyah-Singer index formula for the Dolbeault and Dirac complexes and the
equivariant index formula (\ref{PEeq:ASindex}) can be generalized to a
generic situation of an equivariant index of transversally elliptic
complex (\ref{PEeq:complex}). 

Let $X$ be a real manifold. Let $\pi: T^{*} X \to X$ be the cotangent
bundle. Let $\{E^\bullet \}$ be an indexed set of vector
bundles on $X$ and $\pi^{*} E^\bullet$ be the vector bundles over $T^{*}X$
defined by the pullback. 

The symbol $\sigma(D)$ of a differential operator $D: \Gamma(E) \to \Gamma(F)$ 
(\ref{PEeq:complex}) is a linear operator $\sigma(D): \pi^{*} E \to
\pi^{*} F$ which is defined by taking the highest degree part of the
differential operator and replacing all derivatives $\frac{\partial}{\partial
  x^{\mu}}$ by the conjugate coordinates $p^{\mu}$ in the fibers of
$T^{*} X$. 

For example, for the Laplacian $\Delta: \Omega^{0}(X, \BR) \to \Omega^{0} (X,\BR)$ 
 with highest degree part in some
coordinate system $\{x^{\mu}\}$ given by 
$\Delta =  g^{\mu \nu} \partial_{\mu} \partial_{\nu}$ where $g^{\mu \nu}$ is the inverse Riemannian metric, the symbol of $\Delta$ is a $\Hom(\BR,
\BR)$-valued (i.e. number valued) function on $T^{*}X$ given by 
\begin{equation}
 \sigma(\Delta)  = g^{\mu \nu} p_{\mu} p_{\nu}
\end{equation}
where $p_{\mu}$ are conjugate coordinates (momenta) on the fibers of
$T^{*} X$. 

A differential operator $D: \Gamma(E) \to \Gamma(F)
$ is elliptic if its symbol $\sigma(D): \pi^{*} E \to \pi^{*} F $ is
an isomorphism of vector bundles $\pi^{*}E $ and $\pi^{*} F$ on $T^{*} X$  outside of the
zero section $X \subset T^{*}X $. 

The index of a differential operator $D$ depends only on the topological
class of its symbol in the topological K-theory of vector bundles on
$T^{*} X$.  The Atiyah-Singer formula for the index of the complex (\ref{PEeq:complex})
is\begin{equation}
\boxed{  \ind_{G}(D, X) =\frac{1}{(2\pi)^{\dim_{\BR} X}} \int_{T^{*}X} 
\hat A_{G} (\pi^{*} T_{X}) \ch_{G}(\pi^{* }E^{\bullet})}
\end{equation}

Here $T^{*} X$ denotes the total space of the cotangent bundle of $X$
with canonical orientation such that  $dx^1 \wedge dp_{1}
\wedge dx^2 \wedge dp_2 \dots $ is a positive element of
$\Lambda^{\mathrm{top}} (T^{*} X)$. 

Let $n = \dim_{\BR} X$. Let $\pi^{*} T_{X}$ denote the vector bundle of
dimension $n$ over the total  $T^{*} X$ obtained as pullback of $T_{X}
\to X$ to $T^{*} X$. The $\hat A_{G}$-character of $\pi^{*} T_{X}$ is 
\begin{equation}
\label{PEeq:ASgeneric}
  \hat A_{G} (\pi^{*} T_{X}) = \det{}_{\pi^{*} T_{X}} \left( \frac{ R_{G}}{e^{R_{G}/2}  -
      e^{-R_{G}/2}} \right)
\end{equation}
where $R_{G}$ denotes the $G$-equivariant curvature of the bundle $\pi^{*} T_{X}$. Notice
that the argument of $\hat A$ is $n \times n$ matrix where $n
=\dim_{\BR} T_{X}$ (real dimension of $X$) while if general index formula is specialized to Dirac
operator on Kahler manifold $X$ as in (\ref{PEeq:equivariant-Dirac}) the
argument of the $\hat A$-character is an $n \times n$ matrix where $n = \dim_{\BC}
T_{X}^{1,0}$ (complex dimension of $X$). 

Even though the integration domain $T^{*}X$ is non-compact the integral 
(\ref{PEeq:ASgeneric}) is well-defined because of the ($G$-transversal)
ellipticity of the complex $ \pi^{*} E$. 

For illustration take the complex to be $E_{0} \stackrel{D}{\to} E_1$. 
Since $\sigma(D): \pi^{*} E_0 \to \pi^{*} E_{1}$ is an isomorphism outside
of the zero section we can pick a smooth connection on $\pi^* E_{0}$ and $\pi^* E_1$
such that its curvature on $E_0$ is equal to the curvature on $E_1$ 
away from a compact tubular neighborhood $U_{\epsilon} X$ 
of $X \subset T^* X$. Then  $\ch_{G} (\pi^* E^{\bullet})$
is explicitly vanishing away from $U_{\epsilon} X$ and the integration
over $T^{*}X$ reduces to integration over the compact domain $U_{\epsilon}
X$. 

It is clear that under localization to the fixed points of the $G$-action on
$X$ the general formula (\ref{PEeq:ASgeneric}) reduces to the fixed point
formula (\ref{PEeq:ASindex}). This is due to the fact that the numerator
in the $\hat A$-character $\det_{\pi^{*}  T_{X}} R_{G} =
\Pf_{T_{T_{X}^{*}}} (R_{G})$ is the Euler class of the tangent bundle $T_{T_{X}^{*}}$ to
$T^{*} X$ which cancels with the denominator in (\ref{PEeq:ABlocalization}),
while the restriction of the denominator of (\ref{PEeq:ASgeneric}) to fixed
points is equal to (\ref{PEeq:ASgeneric}) or (\ref{PEeq:ASindex}), because $\det
e^{R_{G}} = 1$, since $R_{G}$ is a curvature of the tangent bundle
$T_{X}$ with orthogonal structure group.

\section{Equivariant cohomological field theories}
Certain field theories first have been interpreted as cohomological and
topological field theories by Witten, see \cite{PEWitten:1990bs},
\cite{PEWitten:1988ze}.

Often the path integral for supersymmetric field theories can be
represented in the form 
\begin{equation}
  Z = \int_{X} \alpha
\end{equation}
where $X$ is the superspace (usually of infinite dimension) of all fields
of the theory.  Moreover, the integrand measure $\alpha$ is closed with
respect to an odd operator  $\delta$ which is typically constructed as a sum
of a supersymmetry algebra generator and a BRST charge
 \begin{equation}
   \delta \alpha = 0
 \end{equation}
The integrand is typically a product  of an exponentiated action
functional $S$, perhaps with insertion of a non-exponentiated observable $\CalO$
\begin{equation}
  \alpha = e^{-S} \mathcal{O}
\end{equation}
so that both $S$ and $\CalO$ are $\delta$-closed
\begin{equation}
  \delta S = 0,  \qquad \delta \mathcal{O} = 0.
\end{equation}

If $X$ is a supermanifold, such as a total space $\Pi E$ of a vector
bundle $E$ (over a base $Y$) with parity inversed  fibers, the equivariant Euler
characteristic class (Pfaffian) in the Atiyah-Bott formula (\ref{PEeq:ABlocalization}) is
replaced by the graded (super) version of the Pfaffian. The weights
associated to  fermionic components contribute inversely compared to the
weights associated to  bosonic components.

Typically, in quantum field theories the base $Y$ of the bundle $E \to
Y$ is the space of
fields. Certain differential  equations (like BPS equations)
are represented by a section $s: Y \to E$. The zero set of the section
 $s^{-1}(0) \subset Y$ are the field configurations which solve the
 equations. For example, in topological self-dual Yang-Mills theory
 (Donaldson-Witten theory) the space $Y$ is the infinite-dimensional
affine space   of all connections on a principal $G$-bundle on a smooth four-manifold
 $M_{4}$. In a given framing,  connections are represented by
 adjoint-valued 1-forms on $M_4$, so $Y \simeq \Omega^{1}(M_4) \otimes
 \ad \mathfrak{g}$. 
 A fiber of the vector bundle $E$ at a given connection $A$ on
 the $G$-bundle on $M_4$  is the space of
 adjoint-valued two-forms  $\Omega^{2+}(M_{4}) \otimes \ad
 \mathfrak{g}$. The section $s: \Omega^{1}(M_4) \otimes \ad \mathfrak{g}
 \to \Omega_{2} $ is represented by the self-dual part of the curvature
 form 
 \begin{equation}
   A \mapsto F_{A}^{+}
 \end{equation}
 The  zeroes of the section $s=0$ are connections $A$ 
that are solutions of the equation $F^{+}_{A}=0$.
The integrand $\alpha$ is the Mathai-Quillen representative of
the Thom class for the bundle $E \to Y$ like in (\ref{PEeq:alphadef}) and
(\ref{PEeq:Vsection}).   The integral over the space of all fields $X = \Pi E$
localizes to the integral over the zeroes $s^{-1}(0)$ of the section , which
in the Donaldson-Witten example is the moduli space of self-dual
connections, called \emph{instanton moduli space}. 

The functional integral version of the localization formula of
Atiyah-Bott has the same formal form 
  \begin{equation}
\label{PEeq:ABlocalizationField}
\boxed{
  \int_{X} \alpha = \int_{F} \frac{ f^{*} \alpha}{\euler (\nu_F)}
}
\end{equation}
except that in the quantum field theory version the space $X$ is
an infinite-dimensional superspace of fields. The $F$ denotes the localization
locus in the space of fields. Let $\Phi_{F} \subset H^{\bullet}(X)$ be
the Poincar\'e dual class to $F$, or Thom class of the inclusion $f: F  \hookrightarrow X$ which provides the isomorphism 
\begin{equation}
  f_{*}: H^\bullet(F) \to H^\bullet(X)
\end{equation}
\begin{equation}
  f_{*}: 1 \mapsto \Phi_{F}
\end{equation}
Let $\nu_{F}$ be the normal bundle to $F$ in $X$. In quantum feld theory
language the space $F$ is called the moduli space or localization locus, 
and  $\nu_F$ is the space of linearized fluctuations of fields
transversal to the localization locus. The  cohomology class of $f^{*}
\Phi_F$ in $H^{\bullet}(F)$ is equal to the Euler class of the normal
bundle $\nu_{F}$
\begin{equation}
  [f^{*} \Phi_F] = e(\nu_F)
\end{equation}

The localization  (\ref{PEeq:ABlocalization}) from $X$ to $F$ exists
whenever the locus $F$ is such that there exists an inverse to the Euler
class $e(\nu_F)$ of its normal bundle in $X$. Two examples of such $F$
have been considered above: 

(i) if $X = \Pi E$ is the total space of a vector
bundle $E \to Y$ with parity inversed fibers, then $F \subset Y \subset
X$ can be taken to
be the set of zeroes $F = s^{-}(0)$ of a generic section $s: Y \to E$

(ii) If $X$ is a $G$-manifold for a compact group $G$, then $F$ can be
taken to be  $F = X^{G}$, the set  of $G$-fixed points on $X$

The formula (\ref{PEeq:ABlocalizationField}) is more general than these
examples.  In practice, in quantum field theory problems, the
localization locus $F$ is found by deforming the form $\alpha$ to 
\begin{equation}
  \alpha_{t} = \alpha \exp( - t \delta V)
\end{equation}
Here $t \in \BR$ is a deformation parameter, and
  $V$ is a fermionic functional on the space of fields, such that $\delta V$ has a trivial
  cohomology class  (the cohomology class $\delta V$ is automatically trivial on effectively compact spaces,
  but on a non-compact space of fields, which usually appears in quantum
  field theory path integrals, one has to take extra care of the
  contributions from the boundary at infinity to ensure that $\delta V$
  has trivial cohomology class).  

If the even part of the functional $\delta V$ is positive definite, then
by sending the paratemeter $t \to \infty$ we can see that the integral 
\begin{equation}
  \int_{X} \alpha \exp(-t \delta V)
\end{equation}
localizes to the locus $F \subset X$ where $\delta V$ vanishes. Such
locus $F$ has an invertible Euler class of its normal bundle in $X$ and
the localization formula (\ref{PEeq:ABlocalizationField}) holds.

In some quantum field theory problems, a compact Lie group $G$ acts on $X$ and $\delta$ is
isomorphic to an equivariant de Rham differential in the Cartan model of
$G$-equivariant cohomology of $X$, so that an element
$\mathbf{a}$ of the
Lie algeba of $G$ appears as a parameter of the partition function $Z$. 

Then the partition function $Z(\mathbf{a})$ can be interpreted as an
element of $H^{\bullet}_{\mathbf{G}}(pt)$, and the Atiyah-Bott localization formula can
be applied to compute $Z(\mathfrak{\mathbf{a}})$. 

There are  are two types of equivariant partition functions. 

In the partition functions of the first type $Z(\mathbf{a})$,  the variable $\mathbf{a}$
is a \emph{parameter} of the quantum field theory such as a coupling
constant, a background field, a choice of vacuum, an asymptotics of fields or
a boundary condition. Such a partition function is typical for a quantum
field theory on a non-compact space, such as the Nekrasov partition function of equivariant gauge theory on
$\BR^{4}_{\ep_1, \ep_2}$ \cite{PENekrasov:2002qd}.

In the partition function of the second type, the variable $\mathbf{a}$
is actually a \emph{dynamical field} of the quantum field theory, so
that the complete partition function is defined by  integration of the
partial partition function $\tilde Z(\mathbf{a}) \in H^{\bullet}_{G}(pt)$
\begin{equation}
  Z = \int_{\mathbf{a} \in \mathbf{g}} \mu(\mathbf{a}) \tilde Z(\mathbf{a})
\end{equation}
where $\mu(a)$ is a certain adjoint invariant volume form on the Lie
algebra $\mathfrak{g}$. The partition function $Z$ of second type is
typical for quantum field theories on compact space-times reviewed 
in \ifvolume{this volume}{\cite{ContributionSummary}}, such as
 the partition function of a supersymmetric gauge theory on $S^4$
 \cite{PEPestun:2007rz} reviewed in \volcite{HO}, 
or on spheres of other dimensions, see summary of results in \volcite{PZ}.

 \section*{Acknowledgements}
 The author is grateful to Bruno Le Floch for a careful proofreading and
 comments on  \ifvolume{this chapter}{this contribution to
   \cite{ContributionSummary}}. The author is supported by ERC QUASIFT grant.

\documentfinish
